\documentclass[reviewcopy]{elsarticle}

\usepackage[reviewcopy]{adndt}
\usepackage{longtable}

\usepackage{amsmath}
\usepackage{amssymb}

\biboptions{square,sort&compress}
\bibpunct[]{[}{]}{,}{n}{}{;}
\begin{document}

\begin{frontmatter}

\journal{Atomic Data and Nuclear Data Tables}

\title{Level density of the $sd$-nuclei $-$ statistical shell-model predictions}

  \author[One]{S. Karampagia\corref{cor1}}
  \ead{E-mail: karampag@nscl.msu.edu}

  \author[Two]{R. A. Senkov}

  \author[One,Three]{V. Zelevinsky}

  \cortext[cor1]{Corresponding author.}

  \address[One]{National Superconducting Cyclotron Laboratory, Michigan State University, East Lansing, MI 48824-1321, USA}

  \address[Two]{Natural Sciences Department, La Guardia Community College, Long Island City, New York 11101, USA}

  \address[Three]{Department of Physics and Astronomy, Michigan State University, East Lansing, MI 48824-1321, USA}

\date{06.13.2017} 

\begin{abstract}

Accurate knowledge of the nuclear level density is important both from a theoretical viewpoint as a powerful instrument for studying nuclear structure and for numerous applications. For example, astrophysical reactions responsible for the nucleosynthesis in the universe can be understood only if we know the nuclear level density.
We use the configuration-interaction nuclear shell model to predict nuclear level density for all nuclei in the $sd$-shell, both total and for individual spins (only with positive parity).
To avoid the diagonalization in large model spaces we use the moments method based on statistical properties of nuclear many-body systems. In the cases where the diagonalization is possible, the results of the moments method practically coincide with those from the shell-model calculations.
Using the computed level densities, we fit the parameters of the Constant Temperature phenomenological model, which can be used by practitioners in their studies of nuclear reactions at excitation energies appropriate for the $sd$-shell nuclei.
\end{abstract}

\end{frontmatter}




\newpage

\tableofcontents
\listofDtables
\listofDfigures
\vskip5pc


\section{Introduction}

The level density as a function of energy and other quantum numbers is an important
characteristic of any realistic quantum system, especially in many-body physics.
The primary use of the level density is in determining the reaction of the system to
an external perturbation, dynamical or thermal. It is hard to overestimate the role
of the nuclear level density as a necessary input to calculations of nuclear reaction
rates, which in their turn, are the paramount ingredients for cross-section calculations
in a broad range of practical problems, extending from nuclear experiments to medical
and technological applications and nucleosynthesis studies.

The many-body level density grows very fast with excitation energy reflecting the
exponential trend of the number of possible excitations of the system compatible with
quantum statistics. The simplest way of theoretical evaluating the level density for
complex nuclei comes from the combinatorial counting of all possible many-nucleon
degrees of freedom which become accessible with increase of available energy.
In the context of the simplest non-interacting Fermi-gas model this was presented in
the classical work by Bethe \cite{Bethe1}. The nuclear level density derived in this way
has the form $\rho(E) \propto e^{\sqrt{aE}}$, which however does not describe well
the low-energy experimental data.

In 1959 Ericson \cite{Ericson} suggested that the use of the level density in the
exponential form, $\rho(E) \propto e^{E/T}$, describes better the experimental data.
The phenomenologically introduced parameter of temperature $T$, is, from a more
modern point of view, the reflection of the effective interaction of the Fermi-liquid
type (in distinction to a perfect Fermi-gas). Following Ericson, the authors of
\cite{CT1, CT2} found that the simple formula
\begin{equation}
\rho(E) = \,\frac{1}{T}\,  e^{(E-E_0)/T}                     \label{0}
\end{equation}
describes well the available experimental data from the low-energy region up to a
certain excitation energy for heavy nuclei and up to neutron or proton resonances and
beyond for light nuclei.

Earlier attempts to account for shell, pairing and deformation effects consisted mainly
in modifying the parameters in the known level density formulas \cite{Ericson, Dilg}
leading also to the widely used back-shifted Fermi-gas (BSFG) model \cite{CT2, Huiz}.
The first analytic way of taking into account pairing forces used the framework of
the BCS superconductivity theory. This effort \cite{Lang, Sano, Decowski} provided better
agreement with experimental data. The pairing force was also included in framework of 
the shell-model level density \cite{Grover, Hillman} and in other
microscopic theories with various single-particle level schemes \cite{Huiz1}.
Soon it was suggested that in order to account for collective, rotational or vibrational,
states shifted to lower excitation energy, the corresponding contributions could be simply
added to the particle or quasiparticle level density (the so-called collective enhancement).
Usually this is accomplished by introducing additional phenomenological
model-dependent parameters \cite{Bjornholm, Ignatyuk,Hansen}.

Monte Carlo methods were also employed providing the level densities with their spin and
parity dependence \cite{Cerf1, Cerf2, Cerf3} and accounting for shell and pairing effects.
Based on a more realistic single-particle level density at the Fermi energy,
global microscopic statistical calculations were proposed in terms of the Extended
Thomas-Fermi plus Strutinsky Integral model \cite{Goriely} and later using the deformed
Hartree-Fock-BCS approach \cite{Demetriou}. These statistical calculations could not
describe the parity dependence of the nuclear level density and its non-statistical
nature at low excitation energies. New microscopic level densities were calculated
using a combinatorial model in the framework of the Hartree-Fock-Bogoliubov formulation
\cite{Hilaire, Goriely1}. In general, the level density found in such approaches turns out
to be not a smooth function of excitation energy clearly reflecting the mean-field subshell
structure.

The nuclear shell model using the full interaction Hamiltonian in the truncated orbital
space is one of the most powerful tools for description and understanding of nuclear
properties as evidenced by the good agreement it reveals with available experimental data.
It is therefore natural to apply the shell model to the problem of the nuclear level
density. However, already from medium-mass nuclei, the dimensions of the Hamiltonian
matrices that need to be diagonalized become prohibitively large, rendering any full
computation practically impossible. To overcome these issues, the shell model Monte
Carlo approach was proposed \cite{Lang1, Alha1} which gives good agreement with the
experimental data \cite{Alha2, Alha3, Alha4, Alha5, Alha6}. The method presented here
also directly uses shell-model Hamiltonians with effective interactions, but it employs the
spectral distribution methods based on the broad experience in studies of statistical
properties of energy spectra and wave functions in many-body interacting systems
\cite{big, Kota1, Kota2}. This approach allows one to  calculate the distributions of
eigenvalues using the moments technique \cite{French, Chang, Haq} and avoiding the
enormous diagonalization. In the cases where such diagonalization is possible, the
moments method fully agrees with the exact results. The significant advantage here 
is the full account of all types of interactions, including incoherent collision-like
processes which smooth out the mean-field remnants and make the resulting level density
a simple function of the excitation energy (``thermodynamic function"). More details about the method
will be given in Section II.

It is the purpose of this paper to collect complete calculations of nuclear level
densities for the entire set of $sd$-nuclei using the shell-model Hamiltonian in the
framework of a statistical approach that does not require the full diagonalization
of the Hamiltonian matrix. Among the $sd$-nuclei, those with $N,Z$=8,20 were
excluded from the calculation, due to their proximity to the shell closure.

The level density given by the shell model is practically a bell-shape curve as a function of energy,
with the maximum in the middle due to the restricted orbital space. We are thus providing
the level density up to energy by a few $MeV$ lower than the peak of the curve.
In the case of the $sd$-shell, the full calculation of the spectrum is also feasible using
the shell-model machinery. The main advantage of the moments method compared to the shell model calculation
is the simplicity and speed of calculation producing the full and exact level
density in a few seconds. The true power of the moments method would be evident either in
an extended orbital space or in higher shells, such as the $pf$ and $pfg_{9/2}$, where the
exact shell-model calculation of the level density through complete diagonalization is 
frequently impossible. We have to note that the $sd$-shell calculation gives only states of 
positive parity, thus the results presented in the text correspond to the positive parity case only.

In Section II the moments method is briefly presented,
while in Section III a description is given of the actual version of the shell-model
Hamiltonian used. In Section IV, the calculated level densities are compared
to the full shell-model results, as well as to the low-energy experimental information.
In Section V, the constant temperature model is fitted to the calculated level densities
and the fitting parameters are tabulated for all individual nuclei. The nuclei close
to the $N=Z$ line display the lowest values of the temperature parameter corresponding
to the certain stability of structure. Spin distributions are presented in Section VI
where we see the systematic occurrence of odd-even angular momentum staggering in
even-even nuclei.

\section{Moments method}

Here we give a short explanation of the statistical approach used in our calculations.
More complete description can be found in previous publications \cite{moments}.
The practical algorithm of computations is given in \cite{comp}. The whole approach is
based on the ideas of quantum complexity which leads to the complicated many-body
stationary states with clear statistical regularities close to the properties of
quantum chaos \cite{big,mitchell,Kota1,Kota2}.

The method starts with the choice of the orbital space (a truncated set of spherical
orbitals for protons and neutrons dictated by the nuclei under study). The mean-field
orbitals with their quantum numbers of spin $j$, parity $(-)^{\ell}$ and single-particle energies
$\epsilon(\ell,j)$ are selected along with the complete set of interaction matrix elements.
All parameters are taken without changes from the standard shell-model versions well known by
their practicality. Currently we work with two-body isospin-invariant interactions although 
the generalizations are possible despite being rather involved computationally.
In this article we limit ourselves by the $sd$-shell model which covers many isotopes from
oxygen to argon.

For a given isotope $(Z,N)$, there are many ways of distributing the valence particles over
available orbitals. Each such configuration, called in this context a partition, $p$, contains
$D_{\alpha p}$ many-body states with exact quantum numbers $\alpha$. The states belonging to 
a given partition are spread over some energy region as a result of interactions
inside the partition. For each partition, the statistical average of an operator $\hat{O}$ over
the states  is defined through the corresponding trace,
\begin{equation}
\langle \hat{O}\rangle_{\alpha p}=\,\frac{1}{D_{\alpha p}}\,{\rm Tr}^{\alpha p} \hat{O}. \label{1}
\end{equation}
In particular, the energy centroid of the partition is the first moment of the Hamiltonian,
\begin{equation}
E_{\alpha p}=\,\frac{1}{D_{\alpha p}}\,{\rm Tr}^{\alpha p} \hat{H};       \label{2}
\end{equation}
this comes directly from the diagonal elements of the Hamiltonian matrix.
The second moment of the Hamiltonian,
\begin{equation}
\sigma^{2}_{\alpha p}=\langle H^{2}\rangle_{\alpha p}-E^{2}_{\alpha p}=\frac{1}{D_{\alpha p}}\,{\rm Tr}^{\alpha p} H^2 - E_{\alpha p}^2,  \label{3}
\end{equation}
is determined by the off-diagonal elements of the Hamiltonian matrix including
the interaction between partitions. Again, no diagonalization is required as this
quantity can be read directly from the Hamiltonian matrix. As known from many
studies \cite{mon,brody,wong} the actual distributions are close to the
Gaussians which is a manifestation of quantum complexity and chaotization. Small
corrections from higher statistical moments can be visible only at relatively high energy
that is typically outside of realistic applicability of the shell model.

Finally, the level density $\rho_{\alpha}(E)$  is found as the sum
of Gaussians weighted with their dimensions for all partitions which include states
at given energy and with quantum numbers $\alpha$,
\begin{equation}
\rho(E;\alpha)=\sum_p D_{\alpha p} G_{\alpha p}(E).   \label{4}
\end{equation}
It turns out that the best results are achieved with the finite range Gaussians,
\begin{equation}
G_{\alpha p}(E) = G(E-E_{\alpha p}+E_{g.s.};\sigma_{\alpha p})  \label{5}
\end{equation}
for each partition. Removing unphysical tails, the Gaussians are
cut off at a distance $\sim 2.8 \ \sigma_{\alpha p}$ from the corresponding centroid and then
renormalized. Practically it is convenient to calculate the invariant traces in the $M$-scheme.
When $\rho_{M}(E)$ is computed, the level density  $\rho_{J}(E)$ for certain spin $J$ is found
as the difference of $\rho_{M=J}$ and $\rho_{M=J+1}$. If the orbital space includes cross-shell
transitions, the admixtures of the spurious center-of-mass excitations appear.
They can be removed by simple recurrence relations \cite{recurr} which however is
not actual for our discussion of the $sd$-shell.

\section{Shell model Hamiltonian}

The success of the whole approach depends critically on the accepted shell-model
(configuration interaction) Hamiltonian. For the $sd$-space below the USDB shell-model
Hamiltonian with two-body interactions \cite{Brown} is used. This spin- and isospin-invariant
effective Hamiltonian has in this space 63 two-body matrix elements, whose values can be
found in Tables I and II of \cite{Brown}, see also \cite{big}. For the construction of
the two-body interaction, the renormalized $G$ matrix from the Bonn-A potential was used,
as given in Table 20 of \cite{Gmatrix}. This interaction is a refinement of the USD Hamiltonian
\cite{Wild, Brown2} being based on an updated set of experimental data considering 608 states
in 77 $sd$-shell nuclei. The results are in good agreement with experimental energy levels
and provide the qualitatively reasonable description of transition rates although the problem
of effective charges still remains.

The NuShellX@MSU version of the shell model \cite{Alex} with the USDB interaction \cite{Brown}
is known to describe quite well the spectroscopy of the $sd$ nuclei. The model space consists
of an inert core (here $^{16}$O) and valence nucleons, in the current case occupying
$d_{5/2}, s_{1/2},$ and $d_{3/2}$ orbitals parameterized by their single-particle energies.
As always, such approach can be physically valid only up to some excitation energy when admixtures
of configurations outside the valence space become important. For many of the nuclei considered here,
$pf$-shell contributions appear already at excitation energy 6 $MeV$. However, our region of relatively
low energy is in many cases sufficiently rich to include practically needed nuclear states.
The formal calculations beyond this physical limit are still useful as an instructive example of
the exactly solved quantum many-body problem where one can study quantum chaos and
thermalization in the isolated many-body system \cite{big,borgonovi}. The variations of the
picture under artificial changes of parameters help in defining the roles of individual matrix
elements in various nuclear phenomena, including the shape transitions. Specifically we will try 
to understand the physics of collective enhancement of the low-lying level density \cite{volya,horoi2,horoi3,sofia,sofia2}.

\section{Comparison with low-energy experimental data and shell-model calculations}

\begin{figure*}[ht!]
\centering
\includegraphics[height=100mm]{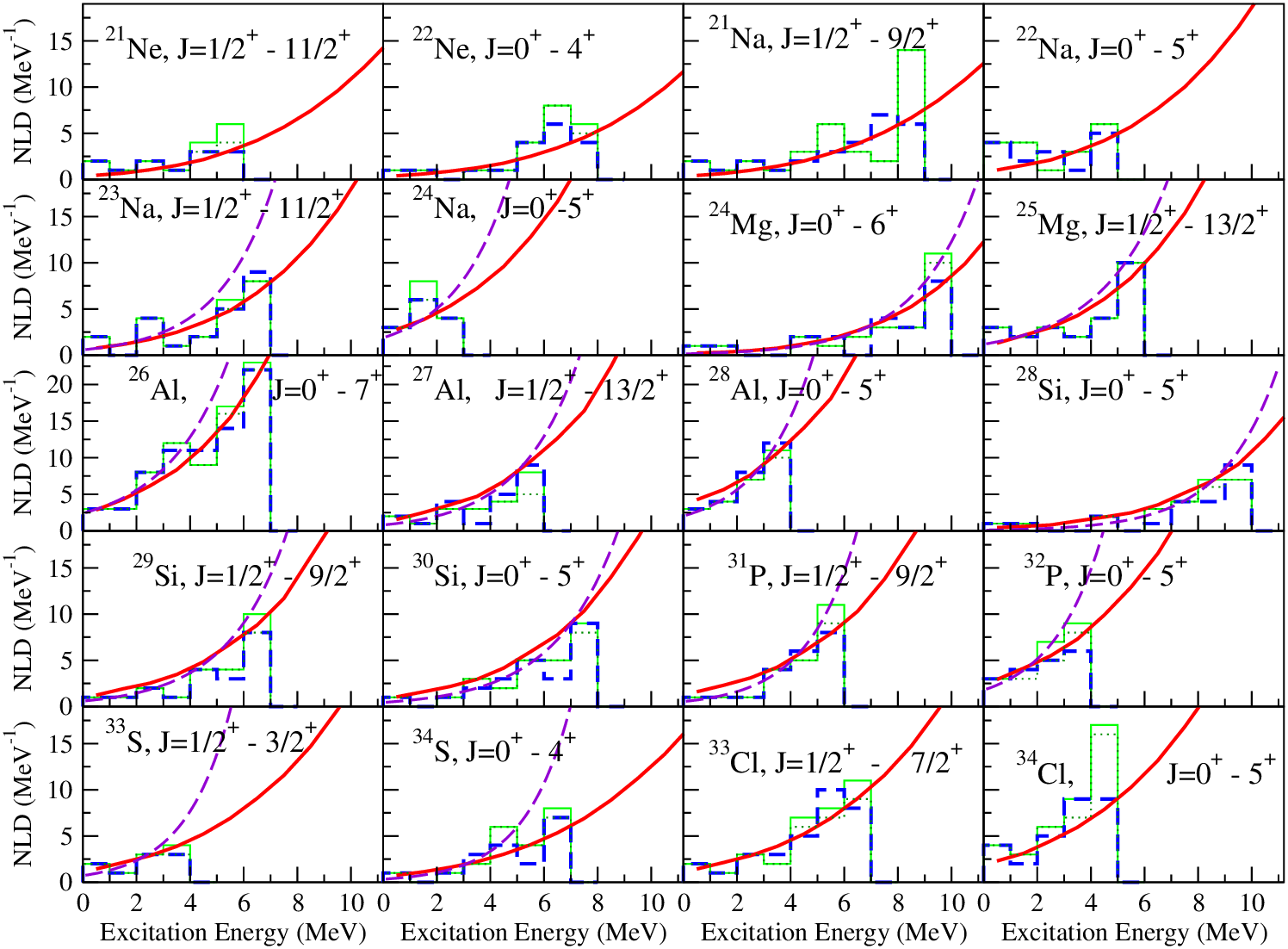}
\caption[]{Comparison of experimental nuclear level densities for positive parity (green and green dotted histograms) 
with the positive parity density calculated with the shell model using the USDB interaction Hamiltonian 
(blue thick dashed histogram) and the moments method (solid red line) for different isotopes of the $sd$-shell. 
The green solid line represents all experimental discrete levels which do not have confirmed negative parity, while the green dotted line 
excludes also the levels of unknown parity. Level densities from both positive and 
negative parity states coming from fitting the constant temperature 
model on the available experimental states \cite{Egidy2} are also presented with a violet dashed line. } \label{fig1}
\end{figure*}

In this section we compare the available experimental low-energy data for nuclei in
the $sd$-shell with the level density derived through the moments method and the shell-model 
calculations, where the USDB two-body interaction \cite{Brown} was used. As seen in 
an example of Fig. \ref{fig1}, the shell model successively describes the low-energy experimental 
levels, while the moments method follows closely the shell model calculations for all nuclei 
where sufficient experimental data are available for comparison in the $sd$-shell. 
However, one can distinguish at least 3 cases where the experimental level density is higher 
than the shell model calculation and by extension, than the moments method; these are the cases 
of the isotopes $^{21}$Na, $^{24}$Mg and $^{34}$Cl. For all three nuclei the level density is 
characteristically high at the very last energy bin and this can be attributed to excitations 
not taken into account by the $sd$-shell model space. Considering $^{21}$Na, which is near the 
beginning of the shell, core excitations coming from the lower subshell is affecting the higher 
excitation energy level density. Similarly, the level density at 5 MeV of 
$^{34}$Cl, a nucleus which is close to the end of the shell, could be enhanced from collective 
excitations coming from the $pf-$shell. As already stated previously, $pf-$shell contributions 
are expected, already at excitation energy of 6 MeV, so the modest surplus of levels of 
$^{24}$Mg at 9 MeV can be explained. 

In the same figure we see the level density produced using the constant temperature parameters 
found from fitting to complete experimental low energy level schemes from Table II of 
\cite{Egidy2}. This level density contains both positive and negative parity states. For these 
nuclei for which constant temperature parameters are provided in \cite{Egidy2}, 
the predicted level density agrees with the level density calculated using the moments method, 
until a certain excitation energy, above which either levels of different parity start appearing, 
or levels of the same parity occur, which cannot be reproduced using the $sd-$shell model space. 

\begin{figure*}[ht!]
\centering
\includegraphics[height=150mm]{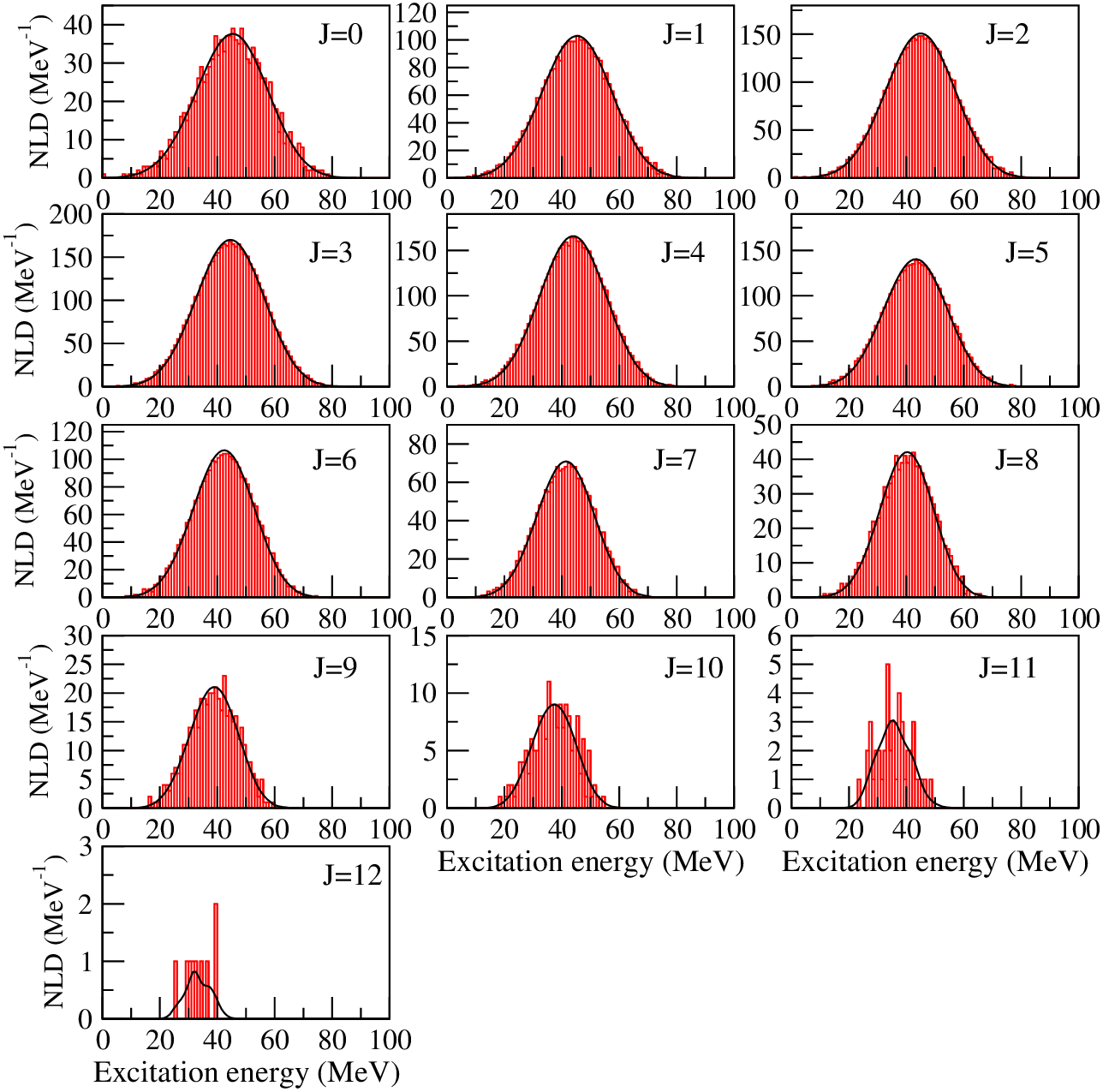}
\caption[]{Comparison of the full nuclear level density calculated for positive parity with the shell model (red histogram) and the moments method (solid black line) for $^{24}$Mg, using the USDB interaction Hamiltonian.}\label{fig2}
\end{figure*}

\begin{figure*}[ht!]
\centering
\includegraphics[height=150mm]{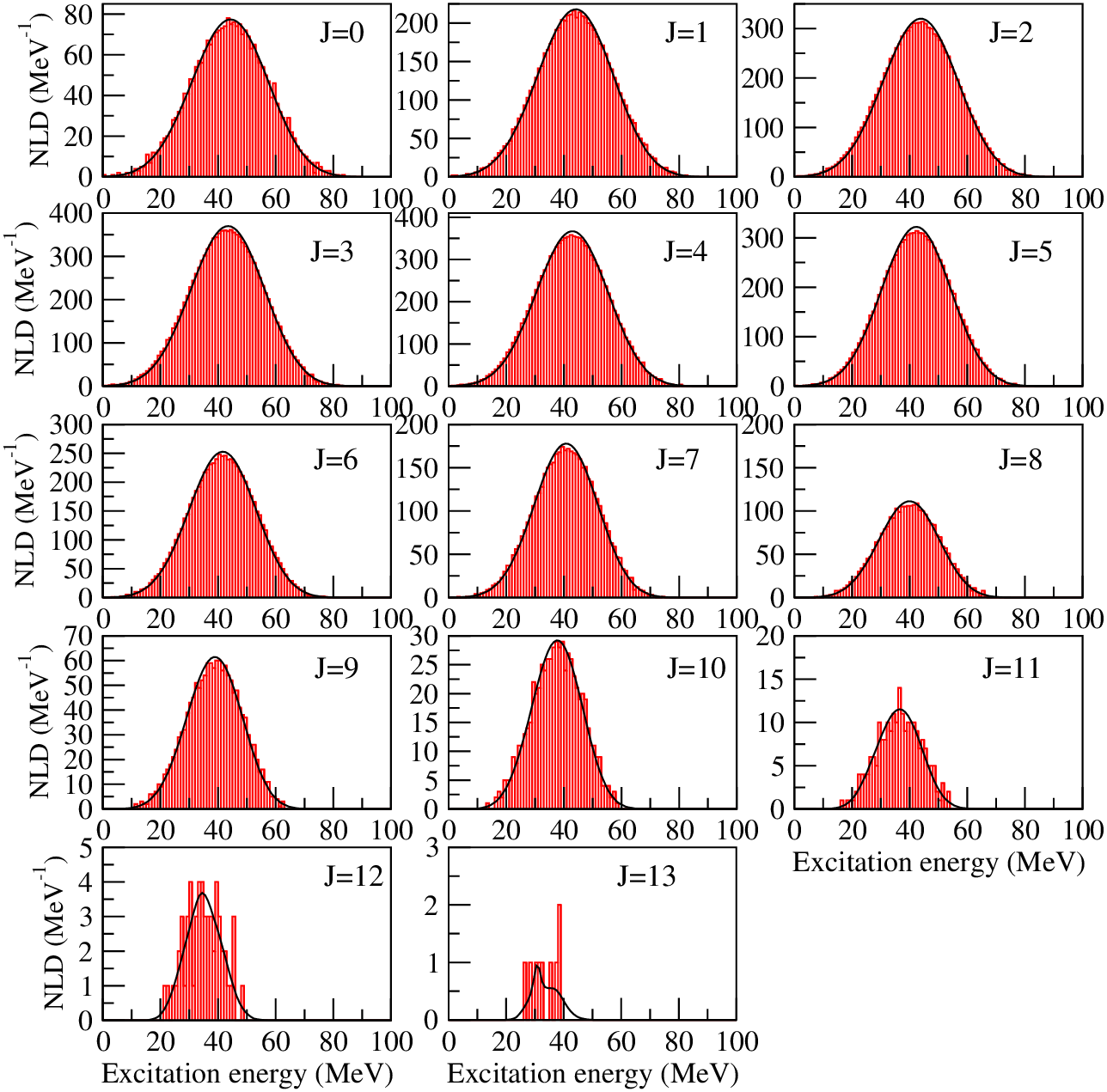}
\caption[]{Comparison of the full nuclear level density calculated for positive parity with the shell model (red histogram) and the moments method (solid black line) for $^{26}$Al, using the USDB interaction Hamiltonian.}\label{fig3}
\end{figure*}

\begin{figure*}[ht!]
\centering
\includegraphics[height=150mm]{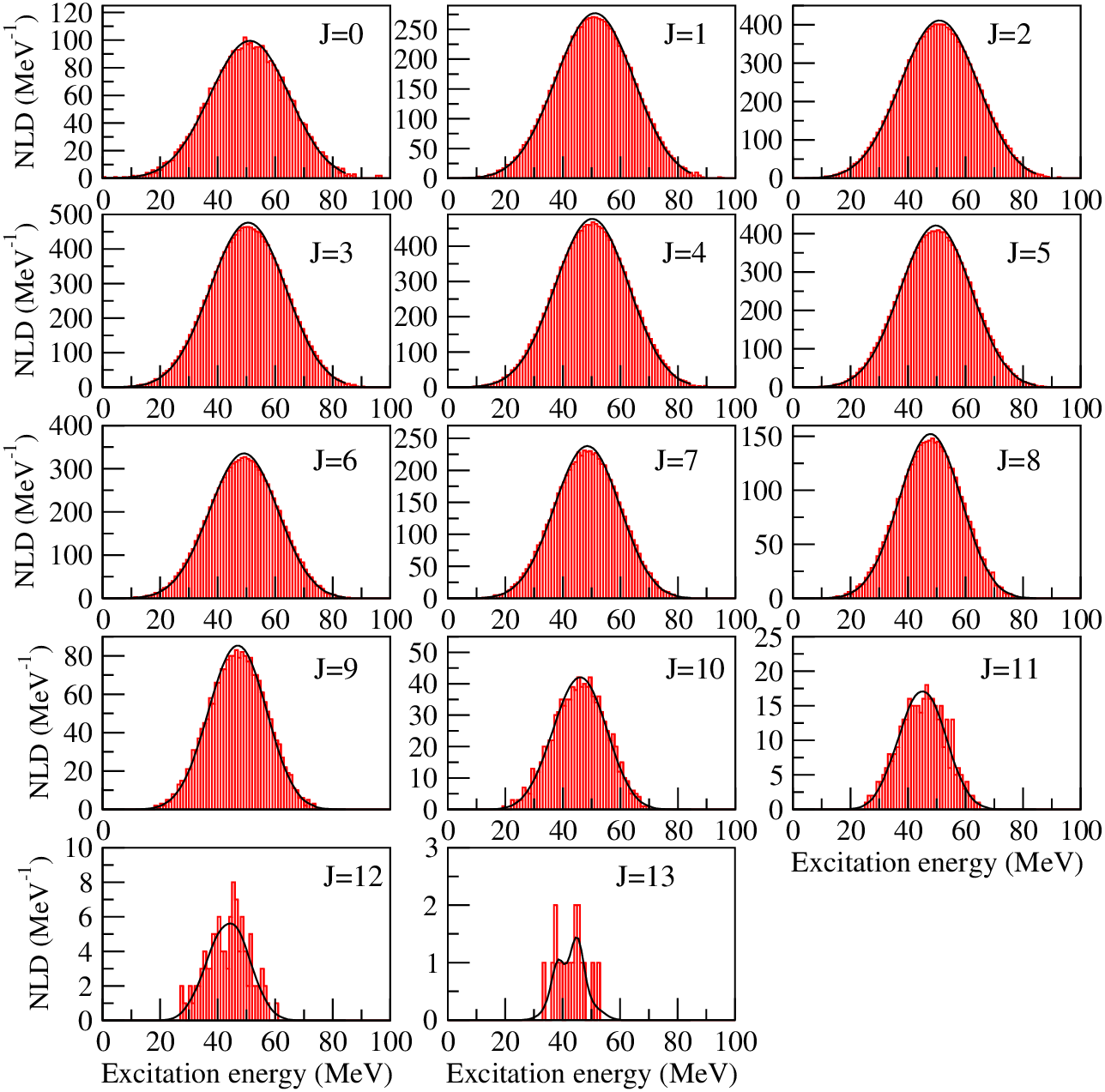}
\caption[]{Comparison of the full nuclear level density calculated for positive parity with the shell model (red histogram) and the moments method (solid black line) for $^{28}$Si, using the USDB interaction Hamiltonian. }\label{fig4}
\end{figure*}

In Figs. \ref{fig2}, \ref{fig3} and \ref{fig4} we compare the full shell-model calculation
represented by the histogram for all spin values of $^{24}$Mg, $^{26}$Al and $^{28}$Si
with the results of the moments method shown by the solid black line.
These selected nuclei are known by the abundance of reliable experimental information.
The agreement between the two methods is excellent for the entire excitation energy region.

\section{Comparison with constant temperature model}

\begin{figure}[ht!]
 \centering
 \includegraphics[height=60mm]{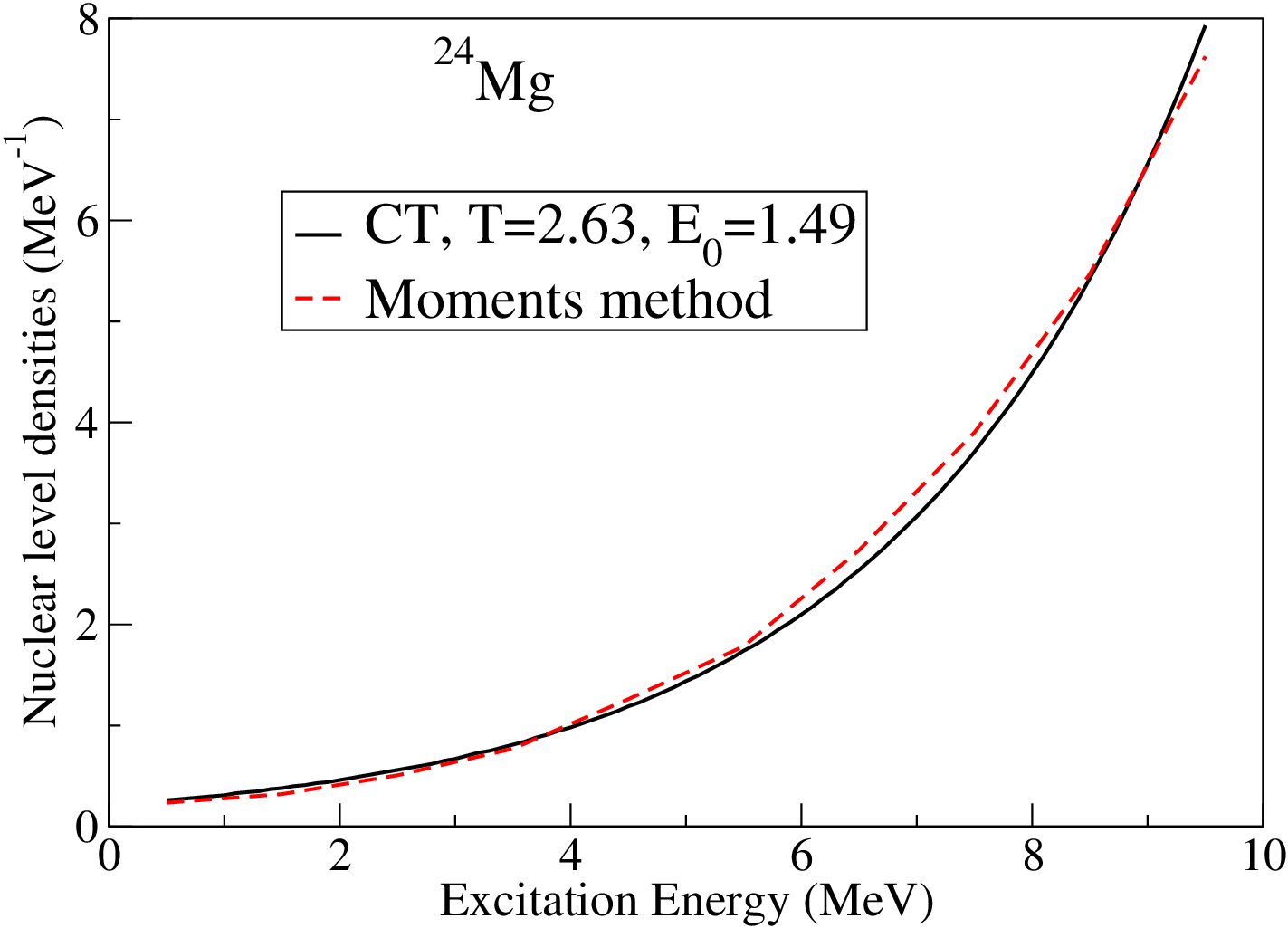}
 \caption{Constant temperature nuclear level density (black solid line), with parameters
 $T$ = 2.63 MeV and $E_0$ = 1.49 MeV found from fitting to the level density of all available $J$ values
 calculated with the moments method (red dashed line), in the excitation energy interval 0$-$10 MeV.}
 \label{fig5}
\end{figure}

In 1965, the authors of \cite{CT1} used different formulas for the cumulative number of levels $N(E)$
up to excitation energy $E$ for light and heavy nuclei \cite{CT2} to describe the available 
experimental data both near the ground state and in the regions of proton or neutron resonances.
It was found that the constant temperature description of the cumulative number $N(E)=e^{(E-E_0)/T}$,
which gives the level density $\rho(E)=N(E)/T$, could satisfactorily fit the experimental data.
Especially for light nuclei, the fit is good over the first 10 MeV or so of excitation energy.
We fit the level density, calculated using the moments method, by the constant temperature formula
given in Eq. (\ref{0}), for all available $J$ values in the interval 0-10 MeV of excitation energy,
for all $sd$-shell nuclei. Plotting the level density calculated from the moments method and
from the constant temperature model with the parameters $E_0$ and $T$ obtained from the fitting, we
see that the two practically coincide in this energy region, an example given in
Fig. \ref{fig5}. The complete set of the parameters obtained from the fit is given in Table
\ref{table1}. It has been shown \cite{moments} that the fit of the moments method level density 
at higher energy by the constant temperature formula leads to a larger value of the parameter $T$, 
in accordance with the idea that the parameter $T$ of our fit can indeed be interpreted as
temperature that grows with increasing excitation energy.

\begin{figure}[ht!]
 \centering
 \includegraphics[height=120mm]{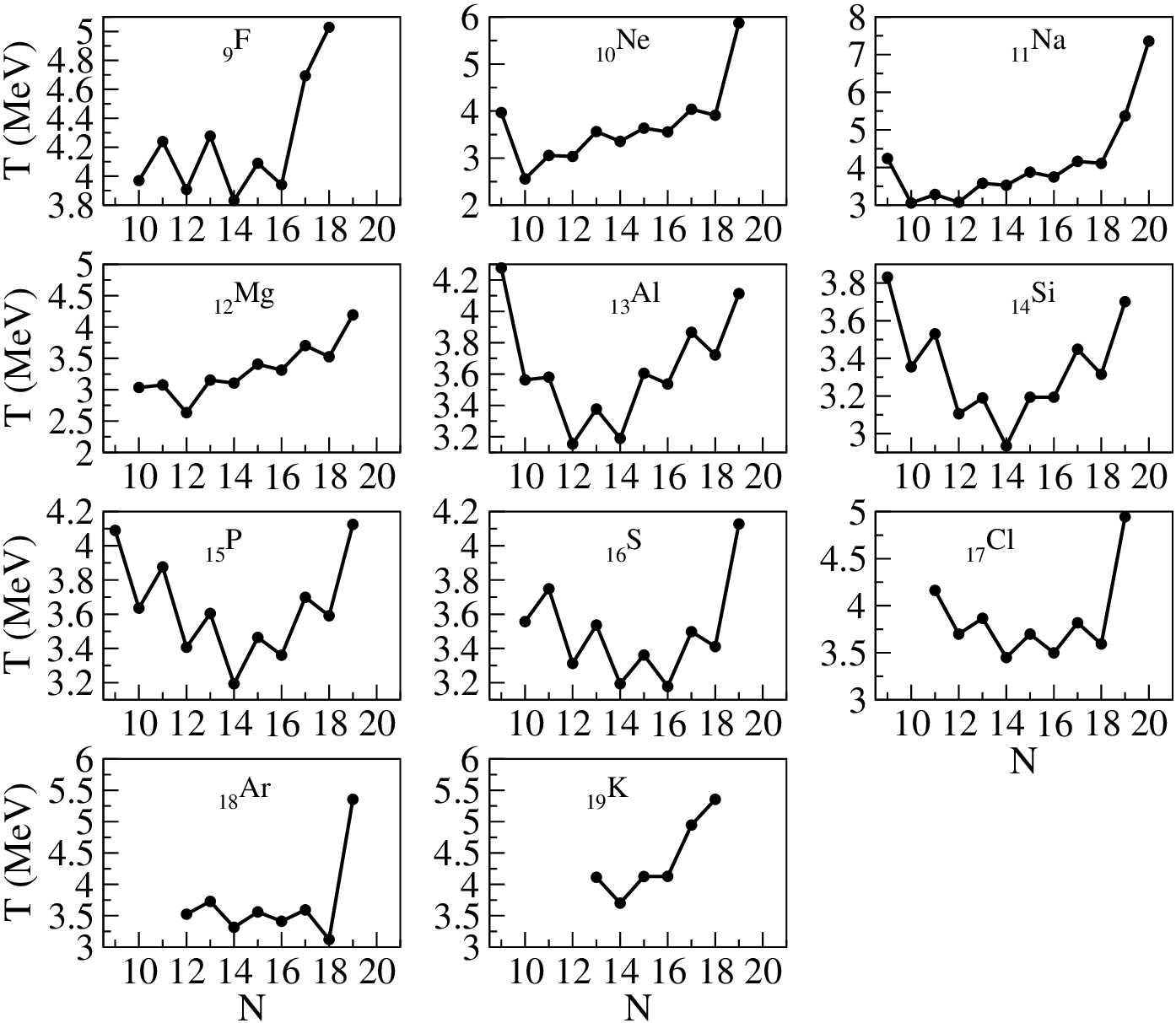}
 \caption{The evolution of the temperature parameter found from fitting the level density of the
 moments method for all available $J$ values to the constant temperature formula, along different isotopic chains 
of the $sd$-shell.}
 \label{fig6}
\end{figure}

While the defined in this way effective temperature is in general constant among the nuclei
considered, having typical values between 3 and 4 MeV, looking at Fig. \ref{fig6} we see a characteristic
behavior of this quantity along the isotopic chains of $sd$-nuclei. First, there is an evident
staggering of the parameter $T$ between even and odd neutron numbers. The temperature
is consistently lower for the isotopes with even number of neutrons compared to the
neighboring isotopes with odd number of neutrons in the same isotopic chain. The
temperature is also higher for the isotopes which are near the closed shells. The parameter $T$
progressively takes smaller values as one moves towards the $N=Z$ or $N=Z\pm 1$ points
for even-even or odd-even nuclei, respectively. For even-even nuclei, the $N=Z$ isotope
is the one with the minimum value of $T$ along an isotopic chain, while for
odd-even nuclei $T$ takes its minimal values for $N=Z\pm 1$ isotopes. These observations
may witness the natural presence of pairing and quartic (alpha-type) correlations making the whole structure
more stable.

In general, the existence of a constant temperature parameter can be understood
as a signature of the gradual process of chaotization of the dynamics that starts already at low
excitation energy and keeps some features of the crossover phase transition \cite{horoi2}
typical for a mesoscopic system. The existence of such a temperature measured by an internal 
thermometer of quasiparticles even in the vicinity of the ground state of an isolated many-body 
system was discussed long ago \cite{MF}, see also \cite{borgonovi}.

\section{Spin distributions}

\begin{figure*}[ht!]
\centering
\includegraphics[height=100mm]{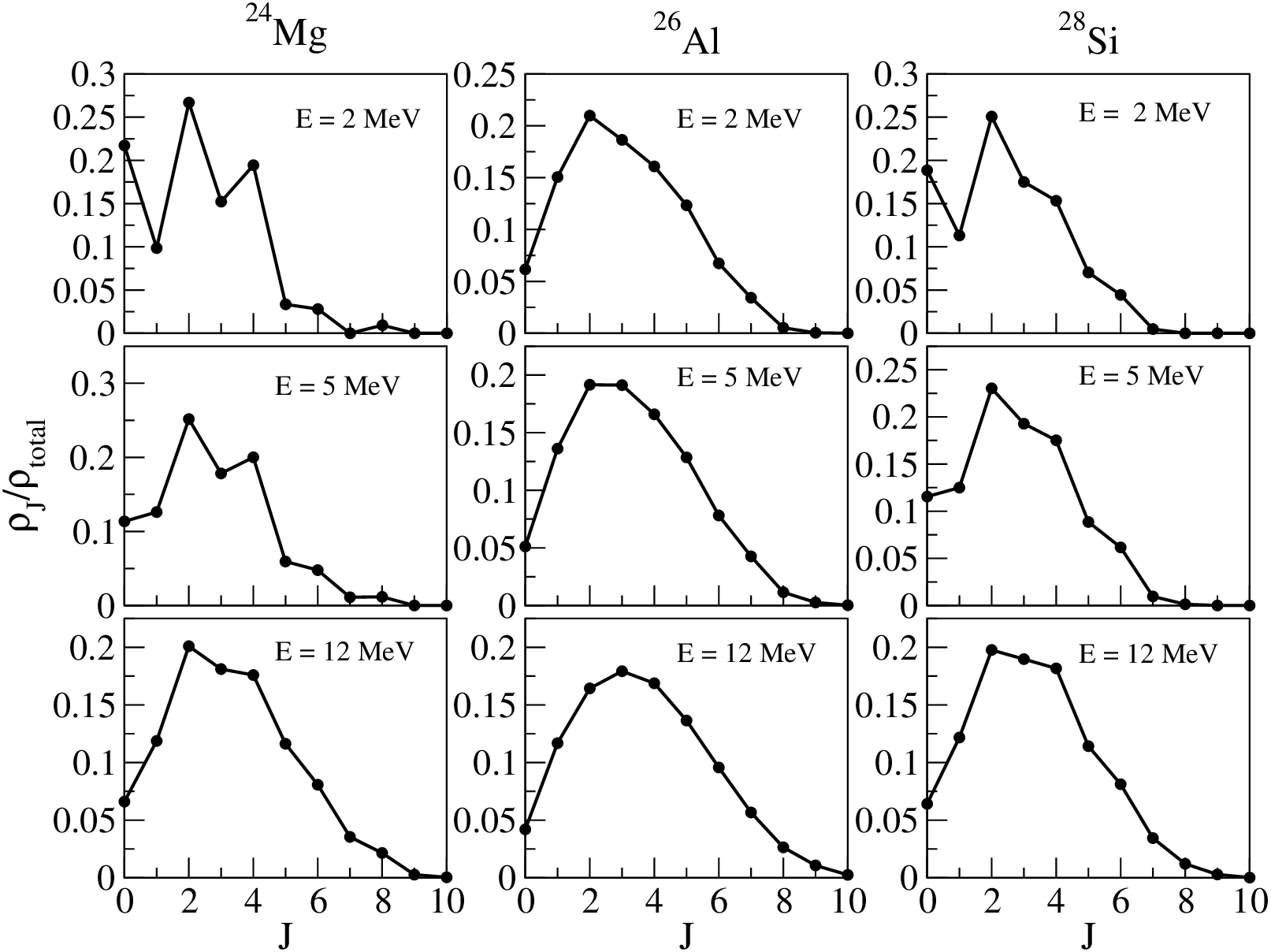}
\caption[]{Spin distribution of the nuclear level density (circles), $\rho_J/\rho_{total}$, at three different excitation energies,  $E=$ 2 MeV, 5 MeV and 12 MeV for $^{24}$Mg, $^{26}$Al, and $^{28}$Si.}
\label{fig7}
\end{figure*}

The spin distribution is equally important for the calculation of nuclear reaction rates
which are frequently sensitive to spin selection rules. The spin distribution
of nuclear levels has been derived microscopically, both in the framework of the shell
model Monte Carlo approach \cite{Alha5} and the static path approximation plus random phase
approximation method \cite{Kaneko}. It was shown that there exists a characteristic
odd-even spin staggering of the level density for even-even nuclei. This phenomenon has been
found also experimentally to occur for all even-even nuclei at low excitation energy
\cite{Egidy1, Egidy2}. In Figure \ref{fig7}, the spin distribution of the nuclear level density, 
$\rho_J/\rho_{total}$, for spin values, $J=0-10$ is plotted for the most studied $^{24}$Mg, $^{26}$Al, 
and $^{28}$Si nuclei at excitation energies 2 MeV, 5 MeV and 12 MeV. Even-even nuclei are distinguished
by the staggering taking place between even and odd values of the angular momentum.
Odd-even angular momentum staggering is also emphasized for smaller excitation energies.\\
\\
\ack

The work on the level density was started with M. Horoi and his students J. Kaiser and M. Ghita. 
The authors are thankful to N. Auerbach, B. A. Brown, A. Renzaglia, and A. Berlaga for numerous discussions. 
The support from the NSF grant PHY-1404442 and the grant No. 2014024 from the 
Binational Science Foundation US-Israel is thankfully acknowledged.

\clearpage

\section*{Figures}

\bigskip

\clearpage

\TableExplanation

In Table \ref{table1}, the $E_0$ and $T$ parameters of the constant temperature
model, for all $sd-$shell nuclei, are tabulated. These are derived from fitting
the constant temperature model formula to the moments method level densities, for
all available $J$ values and for $E=$0-10 MeV.

In Tables 2-108, the moments method level density is tabulated, for all $sd$-shell
nuclei, for all possible $J$ values, up to the excitation energy where the peak
of the level density is reached.

\section*{Table 1. The parameters $E_0$ and $T$ of the constant temperature model,
derived from fitting this formula to the level density calculated using the moments
method, for all available $J$ values, at the energy interval $E=$0-10 MeV, for all
$sd$ isotopes.}

\begin{center}
\begin{tabular}{ll}
$E_0$&   The value of the energy shift of the constant temperature model, derived \\
     &   from its fitting to the moments method level density\\

$T$&    The value of the temperature of the constant temperature model, derived \\
   &    from its fitting to the moments method level density\\
\end{tabular}
\end{center}

\section*{Tables 2-108. Level densities for $sd-$shell nuclei calculated with the moments method using the USDB interaction.}

\begin{center}
\begin{tabular}{ll}
$E$ (MeV)&   Excitation energy in MeV\\

$J$&   The nuclear level density for a specific $J$ value and for specific excitation energy $E$\\
\end{tabular}
\end{center}

\datatables 

\setlength{\tabcolsep}{3.5\tabcolsep}

\renewcommand{\arraystretch}{1.0}

\footnotesize
\begin{longtable}{ r  r  r  r  r  r  r }

\caption{The parameters $E_0$ and $T$ of the constant temperature
model, derived from fitting the constant temperature \\ formula to the level density calculated
using the moments method, for all available $J$ values, at the energy \\ interval $E=$0-10 MeV, for all $sd$ isotopes.}

Nucleus        	&       $E_0$  	&      	$T$    	& &	Nucleus		&     	 $E_0$ 	&      	 $T$   	\\
\hline\\
\endfirsthead\\
\caption[]{(continued)}
Nucleus        	&       $E_0$  	&      	$T$    	& &	Nucleus		&     	 $E_0$ 	&      	 $T$   	\\
\hline\\
\endhead
$^{19}$F	&	-3.42	&	3.97	& &	$^{23}$Si	&	-5.99	&	3.83	\\
$^{20}$F	&	-7.49	&	4.24	& &	$^{24}$Si	&	-3.99	&	3.35	\\
$^{21}$F	&	-6.37	&	3.91	& &	$^{25}$Si	&	-7.96	&	3.53	\\
$^{22}$F	&	-9.95	&	4.28	& &	$^{26}$Si	&	-3.74	&	3.11	\\
$^{23}$F	&	-5.99	&	3.83	& &	$^{27}$Si	&	-5.43	&	3.19	\\
$^{24}$F	&	-7.83	&	4.09	& &	$^{28}$Si	&	-0.61	&	2.94	\\
$^{25}$F	&	-5.40	&	3.94	& &	$^{29}$Si	&	-4.62	&	3.19	\\
$^{26}$F	&	-9.37	&	4.69	& &	$^{30}$Si	&	-3.92	&	3.19	\\
\end{longtable}

\newpage

\setlength{\tabcolsep}{0.3\tabcolsep}

\renewcommand{\arraystretch}{1.0}

\footnotesize
\begin{longtable}{rlllllll}
\caption{Level densities for $^{19}$F}
E $(MeV)$ & J=$\dfrac{1}{2}$ & J=$\dfrac{3}{2}$ & J=$\dfrac{5}{2}$ & J=$\dfrac{7}{2}$ & J=$\dfrac{9}{2}$ & J=$\dfrac{11}{2}$ & J=$\dfrac{13}{2}$ \\
\\
\hline\\
\endhead
0.50 & 0.121 & 0.064 & 0.135 & 0.047 & 0.055 & 0.000 & 0.009 \\
1.50 & 0.172 & 0.103 & 0.204 & 0.080 & 0.091 & 0.000 & 0.029 \\
2.50 & 0.233 & 0.172 & 0.291 & 0.124 & 0.141 & 0.000 & 0.069 \\
3.50 & 0.302 & 0.255 & 0.408 & 0.189 & 0.208 & 0.000 & 0.128 \\
4.50 & 0.371 & 0.366 & 0.541 & 0.291 & 0.288 & 0.029 & 0.192 \\
5.50 & 0.441 & 0.496 & 0.702 & 0.423 & 0.381 & 0.176 & 0.230 \\
6.50 & 0.498 & 0.648 & 0.869 & 0.585 & 0.474 & 0.513 & 0.240 \\
7.50 & 0.546 & 0.818 & 1.039 & 0.765 & 0.572 & 0.734 & 0.239 \\
8.50 & 0.593 & 0.993 & 1.196 & 0.949 & 0.666 & 0.495 & 0.240 \\
9.50 & 0.642 & 1.167 & 1.348 & 1.108 & 0.763 & 0.228 & 0.227 \\
10.50 & 0.696 & 1.335 & 1.479 & 1.238 & 0.862 & 0.207 & 0.185 \\
11.50 & 0.771 & 1.501 & 1.594 & 1.344 & 0.963 & 0.330 & 0.119 \\
12.50 & 0.857 & 1.638 & 1.694 & 1.440 & 1.052 & 0.487 & 0.062 \\
13.50 & 0.945 & 1.749 & 1.780 & 1.516 & 1.110 & 0.600 & 0.025 \\
14.50 & 1.021 & 1.818 & 1.839 & 1.570 & 1.119 & 0.613 & 0.008 \\
\\
\end{longtable}

\setlength{\tabcolsep}{\tabcolsep}

\renewcommand{\arraystretch}{1.0}

\footnotesize
\begin{longtable}{rllllllll}
\caption{Level densities for $^{20}$F}
E $(MeV)$ & J=0 & J=1 & J=2 & J=3 & J=4 & J=5 & J=6 & J=7 \\
\\
\hline\\
\endhead
0.50 & 0.055 & 0.216 & 0.267 & 0.284 & 0.173 & 0.117 & 0.029 & 0.020 \\
1.50 & 0.085 & 0.297 & 0.372 & 0.406 & 0.259 & 0.177 & 0.055 & 0.047 \\
2.50 & 0.123 & 0.412 & 0.523 & 0.564 & 0.356 & 0.253 & 0.095 & 0.084 \\
3.50 & 0.176 & 0.539 & 0.705 & 0.746 & 0.483 & 0.367 & 0.165 & 0.134 \\
4.50 & 0.235 & 0.695 & 0.920 & 0.952 & 0.660 & 0.499 & 0.253 & 0.195 \\
5.50 & 0.309 & 0.865 & 1.172 & 1.196 & 0.859 & 0.659 & 0.354 & 0.250 \\
6.50 & 0.389 & 1.061 & 1.462 & 1.462 & 1.087 & 0.829 & 0.463 & 0.307 \\
7.50 & 0.476 & 1.274 & 1.778 & 1.754 & 1.345 & 1.023 & 0.580 & 0.353 \\
8.50 & 0.571 & 1.498 & 2.101 & 2.046 & 1.653 & 1.223 & 0.694 & 0.400 \\
9.50 & 0.665 & 1.735 & 2.447 & 2.356 & 1.968 & 1.437 & 0.814 & 0.455 \\
10.50 & 0.762 & 1.975 & 2.788 & 2.685 & 2.302 & 1.659 & 0.939 & 0.509 \\
11.50 & 0.853 & 2.213 & 3.120 & 3.012 & 2.627 & 1.873 & 1.058 & 0.547 \\
12.50 & 0.938 & 2.444 & 3.431 & 3.337 & 2.948 & 2.075 & 1.166 & 0.542 \\
13.50 & 1.013 & 2.657 & 3.707 & 3.652 & 3.221 & 2.251 & 1.253 & 0.502 \\
14.50 & 1.076 & 2.850 & 3.938 & 3.941 & 3.444 & 2.386 & 1.306 & 0.478 \\
15.50 & 1.125 & 3.021 & 4.136 & 4.186 & 3.608 & 2.465 & 1.316 & 0.524 \\
16.50 & 1.161 & 3.157 & 4.263 & 4.384 & 3.695 & 2.483 & 1.287 & 0.588 \\
\\
\end{longtable}

\newpage

\setlength{\tabcolsep}{\tabcolsep}

\renewcommand{\arraystretch}{1.0}

\footnotesize
\begin{longtable}{rlllllllll}
\caption{Level densities for $^{21}$F}
E $(MeV)$ & J=$\dfrac{1}{2}$ & J=$\dfrac{3}{2}$ & J=$\dfrac{5}{2}$ & J=$\dfrac{7}{2}$ & J=$\dfrac{9}{2}$ & J=$\dfrac{11}{2}$ & J=$\dfrac{13}{2}$ & J=$\dfrac{15}{2}$ & J=$\dfrac{17}{2}$ \\
\hline\\
\endhead
0.50 & 0.172 & 0.238 & 0.294 & 0.211 & 0.201 & 0.055 & 0.039 & 0.000 & 0.000 \\
1.50 & 0.241 & 0.340 & 0.398 & 0.299 & 0.288 & 0.085 & 0.067 & 0.004 & 0.000 \\
2.50 & 0.317 & 0.462 & 0.568 & 0.444 & 0.387 & 0.141 & 0.104 & 0.008 & 0.000 \\
3.50 & 0.405 & 0.615 & 0.753 & 0.604 & 0.540 & 0.204 & 0.164 & 0.015 & 0.000 \\
4.50 & 0.516 & 0.808 & 0.987 & 0.807 & 0.700 & 0.295 & 0.233 & 0.027 & 0.000 \\
5.50 & 0.649 & 1.039 & 1.285 & 1.056 & 0.902 & 0.430 & 0.335 & 0.043 & 0.000 \\
6.50 & 0.813 & 1.318 & 1.609 & 1.346 & 1.140 & 0.590 & 0.447 & 0.083 & 0.000 \\
7.50 & 0.996 & 1.635 & 1.979 & 1.715 & 1.444 & 0.793 & 0.574 & 0.131 & 0.007 \\
8.50 & 1.192 & 1.985 & 2.427 & 2.127 & 1.765 & 1.027 & 0.714 & 0.203 & 0.043 \\
9.50 & 1.407 & 2.391 & 2.900 & 2.592 & 2.137 & 1.309 & 0.881 & 0.304 & 0.144 \\
10.50 & 1.644 & 2.817 & 3.410 & 3.096 & 2.560 & 1.627 & 1.065 & 0.429 & 0.274\\
11.50 & 1.904 & 3.277 & 3.960 & 3.638 & 2.994 & 1.979 & 1.245 & 0.562 & 0.332\\
12.50 & 2.173 & 3.766 & 4.532 & 4.208 & 3.452 & 2.338 & 1.427 & 0.670 & 0.443\\
13.50 & 2.451 & 4.255 & 5.112 & 4.788 & 3.925 & 2.698 & 1.604 & 0.745 & 0.523\\
14.50 & 2.738 & 4.750 & 5.708 & 5.389 & 4.384 & 3.048 & 1.769 & 0.774 & 0.218\\
15.50 & 3.022 & 5.244 & 6.270 & 5.951 & 4.832 & 3.381 & 1.915 & 0.795 & 0.074\\
16.50 & 3.283 & 5.705 & 6.804 & 6.486 & 5.232 & 3.668 & 2.031 & 0.840 & 0.236\\
17.50 & 3.518 & 6.116 & 7.266 & 6.947 & 5.586 & 3.896 & 2.108 & 0.897 & 0.402\\
18.50 & 3.716 & 6.467 & 7.652 & 7.328 & 5.865 & 4.060 & 2.145 & 0.934 & 0.251\\
19.50 & 3.868 & 6.746 & 7.949 & 7.618 & 6.064 & 4.135 & 2.125 & 0.921 & 0.057\\
20.50 & 3.975 & 6.946 & 8.147 & 7.802 & 6.175 & 4.124 & 2.067 & 0.882 & 0.000\\
\\
\end{longtable}

\setlength{\tabcolsep}{\tabcolsep}

\renewcommand{\arraystretch}{1.0}

\footnotesize
\begin{longtable}{rllllllllll}
\caption{Level densities for $^{22}$F}
E $(MeV)$ & J=0 & J=1 & J=2 & J=3 & J=4 & J=5 & J=6 & J=7 & J=8 & J=9 \\
\hline\\
\endhead
0.50 & 0.118 & 0.392 & 0.508 & 0.525 & 0.400 & 0.284 & 0.078 & 0.023 & 0.000 & 0.000 \\
1.50 & 0.163 & 0.519 & 0.672 & 0.696 & 0.539 & 0.384 & 0.139 & 0.038 & 0.000 & 0.000 \\
2.50 & 0.227 & 0.670 & 0.891 & 0.910 & 0.717 & 0.509 & 0.208 & 0.068 & 0.000 & 0.000 \\
3.50 & 0.296 & 0.862 & 1.156 & 1.162 & 0.934 & 0.686 & 0.300 & 0.121 & 0.008 & 0.000 \\
4.50 & 0.378 & 1.071 & 1.447 & 1.479 & 1.215 & 0.891 & 0.434 & 0.184 & 0.020 & 0.000 \\
5.50 & 0.486 & 1.353 & 1.861 & 1.896 & 1.534 & 1.151 & 0.595 & 0.268 & 0.057 & 0.000 \\
6.50 & 0.611 & 1.655 & 2.283 & 2.336 & 1.963 & 1.459 & 0.801 & 0.381 & 0.110 & 0.000 \\
7.50 & 0.749 & 1.997 & 2.762 & 2.837 & 2.423 & 1.840 & 1.040 & 0.509 & 0.181 & 0.000 \\
\end{longtable}

\newpage

\setlength{\tabcolsep}{\tabcolsep}

\renewcommand{\arraystretch}{1.0}

\footnotesize
\begin{longtable}{rlllllllll}
\caption{Level densities for $^{23}$F}
E $(MeV)$ & J=$\dfrac{1}{2}$ & J=$\dfrac{3}{2}$ & J=$\dfrac{5}{2}$ & J=$\dfrac{7}{2}$ & J=$\dfrac{9}{2}$ & J=$\dfrac{11}{2}$ & J=$\dfrac{13}{2}$ & J=$\dfrac{15}{2}$ & J=$\dfrac{17}{2}$ \\
\hline\\
\endhead
0.50 & 0.157 & 0.240 & 0.355 & 0.222 & 0.177 & 0.053 & 0.011 & 0.000 & 0.000 \\
1.50 & 0.221 & 0.330 & 0.493 & 0.302 & 0.246 & 0.095 & 0.020 & 0.000 & 0.000 \\
2.50 & 0.294 & 0.434 & 0.628 & 0.442 & 0.353 & 0.141 & 0.046 & 0.000 & 0.000 \\
3.50 & 0.389 & 0.581 & 0.814 & 0.587 & 0.483 & 0.212 & 0.078 & 0.000 & 0.000 \\
4.50 & 0.500 & 0.761 & 1.015 & 0.784 & 0.635 & 0.303 & 0.127 & 0.010 & 0.000 \\
5.50 & 0.647 & 1.015 & 1.336 & 1.040 & 0.860 & 0.414 & 0.204 & 0.021 & 0.003 \\
6.50 & 0.800 & 1.275 & 1.665 & 1.387 & 1.148 & 0.550 & 0.297 & 0.039 & 0.008 \\
7.50 & 0.987 & 1.610 & 2.049 & 1.750 & 1.449 & 0.779 & 0.420 & 0.072 & 0.019 \\
\end{longtable}

\setlength{\tabcolsep}{\tabcolsep}

\renewcommand{\arraystretch}{1.0}

\footnotesize
\begin{longtable}{rllllllllll}
\caption{Level densities for $^{24}$F}
E $(MeV)$ & J=0 & J=1 & J=2 & J=3 & J=4 & J=5 & J=6 & J=7 & J=8 & J=9  \\
\hline\\
\endhead
0.50 & 0.097 & 0.296 & 0.428 & 0.399 & 0.247 & 0.120 & 0.027 & 0.005 & 0.000 & 0.000 \\
1.50 & 0.127 & 0.407 & 0.584 & 0.530 & 0.359 & 0.171 & 0.053 & 0.010 & 0.000 & 0.000 \\
2.50 & 0.185 & 0.520 & 0.748 & 0.714 & 0.478 & 0.257 & 0.085 & 0.023 & 0.000 & 0.000 \\
3.50 & 0.237 & 0.672 & 0.970 & 0.916 & 0.665 & 0.352 & 0.130 & 0.042 & 0.000 & 0.000 \\
4.50 & 0.300 & 0.848 & 1.204 & 1.186 & 0.900 & 0.517 & 0.204 & 0.070 & 0.000 & 0.000 \\
5.50 & 0.374 & 1.046 & 1.493 & 1.495 & 1.153 & 0.692 & 0.299 & 0.108 & 0.010 & 0.000 \\
6.50 & 0.460 & 1.287 & 1.817 & 1.855 & 1.479 & 0.942 & 0.460 & 0.154 & 0.028 & 0.000 \\
7.50 & 0.568 & 1.565 & 2.192 & 2.283 & 1.857 & 1.234 & 0.660 & 0.234 & 0.063 & 0.000 \\
\end{longtable}

\newpage

\setlength{\tabcolsep}{\tabcolsep}

\renewcommand{\arraystretch}{1.0}

\footnotesize
\begin{longtable}{rlllllllll}
\caption{Level densities for $^{25}$F}
E $(MeV)$ & J=$\dfrac{1}{2}$ & J=$\dfrac{3}{2}$ & J=$\dfrac{5}{2}$ & J=$\dfrac{7}{2}$ & J=$\dfrac{9}{2}$ & J=$\dfrac{11}{2}$ & J=$\dfrac{13}{2}$ & J=$\dfrac{15}{2}$ & J=$\dfrac{17}{2}$ \\
\hline\\
\endhead
0.50 & 0.190 & 0.222 & 0.311 & 0.115 & 0.086 & 0.021 & 0.000 & 0.000 & 0.000 \\
1.50 & 0.258 & 0.301 & 0.421 & 0.191 & 0.134 & 0.034 & 0.006 & 0.000 & 0.000 \\
2.50 & 0.325 & 0.418 & 0.536 & 0.282 & 0.216 & 0.051 & 0.015 & 0.000 & 0.000 \\
3.50 & 0.403 & 0.549 & 0.688 & 0.426 & 0.322 & 0.088 & 0.032 & 0.000 & 0.000 \\
4.50 & 0.495 & 0.705 & 0.860 & 0.589 & 0.451 & 0.132 & 0.060 & 0.000 & 0.000 \\
5.50 & 0.594 & 0.882 & 1.061 & 0.797 & 0.606 & 0.191 & 0.095 & 0.000 & 0.000 \\
6.50 & 0.725 & 1.121 & 1.346 & 1.059 & 0.790 & 0.289 & 0.148 & 0.000 & 0.000 \\
7.50 & 0.864 & 1.391 & 1.649 & 1.398 & 1.008 & 0.404 & 0.216 & 0.005 & 0.000 \\
\end{longtable}

\setlength{\tabcolsep}{\tabcolsep}

\renewcommand{\arraystretch}{1.0}

\footnotesize
\begin{longtable}{rllllllll}
\caption{Level densities for $^{26}$F}
E $(MeV)$ & J=0 & J=1 & J=2 & J=3 & J=4 & J=5 & J=6 & J=7 \\
\hline\\
\endhead
0.50 & 0.104 & 0.388 & 0.436 & 0.287 & 0.266 & 0.052 & 0.000 & 0.000 \\
1.50 & 0.137 & 0.489 & 0.566 & 0.386 & 0.330 & 0.095 & 0.000 & 0.000 \\
2.50 & 0.174 & 0.609 & 0.694 & 0.541 & 0.406 & 0.160 & 0.000 & 0.000 \\
3.50 & 0.222 & 0.742 & 0.872 & 0.732 & 0.523 & 0.242 & 0.022 & 0.000 \\
4.50 & 0.269 & 0.878 & 1.054 & 0.959 & 0.687 & 0.341 & 0.067 & 0.000 \\
5.50 & 0.323 & 1.020 & 1.256 & 1.221 & 0.872 & 0.459 & 0.160 & 0.000 \\
6.50 & 0.383 & 1.174 & 1.489 & 1.508 & 1.098 & 0.596 & 0.305 & 0.060 \\
7.50 & 0.450 & 1.341 & 1.741 & 1.812 & 1.371 & 0.744 & 0.441 & 0.274 \\
8.50 & 0.533 & 1.535 & 2.019 & 2.136 & 1.662 & 0.930 & 0.515 & 0.417 \\
9.50 & 0.627 & 1.755 & 2.321 & 2.452 & 1.960 & 1.144 & 0.546 & 0.237 \\
10.50 & 0.732 & 1.977 & 2.623 & 2.761 & 2.237 & 1.386 & 0.583 & 0.130 \\
11.50 & 0.846 & 2.214 & 2.934 & 3.053 & 2.495 & 1.617 & 0.699 & 0.251 \\
12.50 & 0.956 & 2.439 & 3.250 & 3.344 & 2.696 & 1.851 & 0.904 & 0.414 \\
13.50 & 1.058 & 2.646 & 3.557 & 3.604 & 2.861 & 2.027 & 1.151 & 0.520 \\
14.50 & 1.142 & 2.846 & 3.849 & 3.830 & 3.005 & 2.168 & 1.340 & 0.565 \\
15.50 & 1.212 & 3.027 & 4.094 & 4.013 & 3.136 & 2.265 & 1.417 & 0.520 \\
16.50 & 1.264 & 3.184 & 4.278 & 4.144 & 3.260 & 2.328 & 1.356 & 0.357 \\
\\
\end{longtable}

\newpage

\setlength{\tabcolsep}{\tabcolsep}

\renewcommand{\arraystretch}{1.0}

\footnotesize
\begin{longtable}{rlllllll}
\caption{Level densities for $^{27}$F}
E $(MeV)$ & J=$\dfrac{1}{2}$ & J=$\dfrac{3}{2}$ & J=$\dfrac{5}{2}$ & J=$\dfrac{7}{2}$ & J=$\dfrac{9}{2}$ & J=$\dfrac{11}{2}$ & J=$\dfrac{13}{2}$ \\
\hline\\
\endhead
0.50 & 0.219 & 0.201 & 0.227 & 0.112 & 0.091 & 0.000 & 0.000 \\
1.50 & 0.287 & 0.283 & 0.296 & 0.206 & 0.163 & 0.000 & 0.000 \\
2.50 & 0.348 & 0.393 & 0.392 & 0.309 & 0.229 & 0.000 & 0.000 \\
3.50 & 0.396 & 0.522 & 0.557 & 0.374 & 0.269 & 0.000 & 0.000 \\
4.50 & 0.433 & 0.655 & 0.782 & 0.429 & 0.315 & 0.022 & 0.000 \\
5.50 & 0.467 & 0.754 & 1.021 & 0.513 & 0.366 & 0.062 & 0.000 \\
6.50 & 0.505 & 0.823 & 1.198 & 0.648 & 0.460 & 0.140 & 0.000 \\
7.50 & 0.555 & 0.870 & 1.297 & 0.848 & 0.582 & 0.249 & 0.000 \\
8.50 & 0.632 & 0.936 & 1.330 & 1.024 & 0.710 & 0.350 & 0.207 \\
9.50 & 0.712 & 1.053 & 1.386 & 1.154 & 0.793 & 0.389 & 0.764 \\
10.50 & 0.802 & 1.243 & 1.524 & 1.260 & 0.843 & 0.342 & 0.000 \\
11.50 & 0.895 & 1.479 & 1.730 & 1.370 & 0.862 & 0.269 & 0.000 \\
12.50 & 0.997 & 1.717 & 1.913 & 1.477 & 0.892 & 0.293 & 0.000 \\
13.50 & 1.089 & 1.887 & 1.990 & 1.530 & 0.946 & 0.503 & 0.000 \\
\\
\end{longtable}

\newpage

\setlength{\tabcolsep}{\tabcolsep}

\renewcommand{\arraystretch}{1.0}

\footnotesize
\begin{longtable}{rccccccc}
\caption{Level densities for $^{19}$Ne}
E $(MeV)$ & J=$\dfrac{1}{2}$ & J=$\dfrac{3}{2}$ & J=$\dfrac{5}{2}$ & J=$\dfrac{7}{2}$ & J=$\dfrac{9}{2}$ & J=$\dfrac{11}{2}$ & J=$\dfrac{13}{2}$ \\
\hline\\
\endhead
0.50 & 0.121 & 0.064 & 0.135 & 0.047 & 0.055 & 0.000 & 0.009  \\
1.50 & 0.172 & 0.103 & 0.204 & 0.080 & 0.091 & 0.000 & 0.029  \\
2.50 & 0.233 & 0.172 & 0.291 & 0.124 & 0.141 & 0.000 & 0.069  \\
3.50 & 0.302 & 0.255 & 0.408 & 0.189 & 0.208 & 0.000 & 0.128  \\
4.50 & 0.371 & 0.366 & 0.541 & 0.291 & 0.288 & 0.029 & 0.192  \\
5.50 & 0.441 & 0.496 & 0.702 & 0.423 & 0.381 & 0.176 & 0.230  \\
6.50 & 0.498 & 0.648 & 0.869 & 0.585 & 0.474 & 0.513 & 0.240  \\
7.50 & 0.546 & 0.818 & 1.039 & 0.765 & 0.572 & 0.734 & -  \\
8.50 & 0.593 & 0.993 & 1.196 & 0.949 & 0.666 & 0.495 & -  \\
9.50 & 0.642 & 1.167 & 1.348 & 1.108 & 0.763 & 0.228 & -  \\
10.50 & 0.696 & 1.335 & 1.479 & 1.238 & 0.862 & 0.207 & -  \\
11.50 & 0.771 & 1.501 & 1.594 & 1.344 & 0.963 & 0.330 & -  \\
12.50 & 0.857 & 1.638 & 1.694 & 1.440 & 1.052 & 0.487 & -  \\
13.50 & 0.945 & 1.749 & 1.780 & 1.516 & 1.110 & 0.600 & -  \\
14.50 & 1.021 & 1.818 & 1.839 & 1.570 & 1.119 & 0.613 & - \\
\\
\end{longtable}

\setlength{\tabcolsep}{\tabcolsep}

\renewcommand{\arraystretch}{1.0}

\footnotesize
\begin{longtable}{rccccccccc}
\caption{Level densities for $^{20}$Ne}
E $(MeV)$ & J=0 & J=1 & J=2 & J=3 & J=4 & J=5 & J=6 & J=7 & J=8 \\
\hline\\
\endhead
0.50 & 0.035 & 0.000 & 0.023 & 0.000 & 0.012 & 0.000 & 0.000 & 0.000 & 0.000 \\
1.50 & 0.052 & 0.000 & 0.033 & 0.000 & 0.019 & 0.000 & 0.000 & 0.000 & 0.000 \\
2.50 & 0.072 & 0.000 & 0.066 & 0.022 & 0.030 & 0.000 & 0.000 & 0.000 & 0.000 \\
3.50 & 0.098 & 0.005 & 0.098 & 0.036 & 0.044 & 0.000 & 0.000 & 0.000 & 0.000 \\
4.50 & 0.135 & 0.016 & 0.144 & 0.056 & 0.084 & 0.009 & 0.008 & 0.000 & 0.000 \\
5.50 & 0.183 & 0.035 & 0.200 & 0.084 & 0.127 & 0.016 & 0.015 & 0.000 & 0.000 \\
6.50 & 0.234 & 0.055 & 0.297 & 0.122 & 0.205 & 0.042 & 0.039 & 0.000 & 0.000 \\
7.50 & 0.293 & 0.106 & 0.412 & 0.187 & 0.289 & 0.070 & 0.068 & 0.000 & 0.006 \\
\end{longtable}

\newpage

\setlength{\tabcolsep}{\tabcolsep}

\renewcommand{\arraystretch}{1.0}

\footnotesize
\begin{longtable}{rcccccccccc}
\caption{Level densities for $^{21}$Ne}
E $(MeV)$ & J=$\dfrac{1}{2}$ & J=$\dfrac{3}{2}$ & J=$\dfrac{5}{2}$ & J=$\dfrac{7}{2}$ & J=$\dfrac{9}{2}$ & J=$\dfrac{11}{2}$ & J=$\dfrac{13}{2}$ & J=$\dfrac{15}{2}$ & J=$\dfrac{17}{2}$ & J=$\dfrac{19}{2}$ \\
\hline\\
\endhead
0.50 & 0.062 & 0.105 & 0.126 & 0.082 & 0.068 & 0.017 & 0.010 & 0.000 & 0.000 & 0.000 \\
1.50 & 0.101 & 0.147 & 0.175 & 0.130 & 0.102 & 0.030 & 0.017 & 0.000 & 0.000 & 0.000 \\
2.50 & 0.143 & 0.230 & 0.280 & 0.184 & 0.159 & 0.059 & 0.030 & 0.000 & 0.000 & 0.000 \\
3.50 & 0.192 & 0.310 & 0.393 & 0.304 & 0.238 & 0.098 & 0.049 & 0.005 & 0.000 & 0.000 \\
4.50 & 0.269 & 0.431 & 0.526 & 0.426 & 0.353 & 0.157 & 0.085 & 0.015 & 0.007 & 0.000 \\
5.50 & 0.372 & 0.622 & 0.752 & 0.617 & 0.485 & 0.259 & 0.133 & 0.028 & 0.016 & 0.000 \\
6.50 & 0.492 & 0.813 & 1.005 & 0.852 & 0.700 & 0.370 & 0.205 & 0.048 & 0.033 & 0.000 \\
7.50 & 0.643 & 1.075 & 1.329 & 1.123 & 0.955 & 0.546 & 0.313 & 0.086 & 0.068 & 0.000 \\
\end{longtable}

\setlength{\tabcolsep}{\tabcolsep}

\renewcommand{\arraystretch}{1.0}

\footnotesize
\begin{longtable}{rccccccccccc}
\caption{Level densities for $^{22}$Ne}
E $(MeV)$ & J=0 & J=1 & J=2 & J=3 & J=4 & J=5 & J=6 & J=7 & J=8 & J=9 & J=10  \\
\hline\\
\endhead
0.50 & 0.083 & 0.043 & 0.137 & 0.053 & 0.087 & 0.025 & 0.015 & 0.003 & 0.000 & 0.000 & 0.000  \\
1.50 & 0.108 & 0.067 & 0.199 & 0.090 & 0.137 & 0.038 & 0.037 & 0.005 & 0.000 & 0.000 & 0.000  \\
2.50 & 0.139 & 0.121 & 0.294 & 0.144 & 0.200 & 0.056 & 0.061 & 0.008 & 0.002 & 0.000 & 0.000  \\
3.50 & 0.175 & 0.163 & 0.390 & 0.238 & 0.309 & 0.098 & 0.088 & 0.022 & 0.004 & 0.000 & 0.000  \\
4.50 & 0.234 & 0.241 & 0.545 & 0.327 & 0.421 & 0.151 & 0.138 & 0.033 & 0.008 & 0.000 & 0.000  \\
5.50 & 0.313 & 0.357 & 0.751 & 0.518 & 0.581 & 0.245 & 0.216 & 0.054 & 0.019 & 0.000 & 0.000  \\
6.50 & 0.405 & 0.478 & 0.986 & 0.722 & 0.836 & 0.372 & 0.319 & 0.084 & 0.037 & 0.000 & 0.000  \\
7.50 & 0.501 & 0.663 & 1.274 & 0.979 & 1.125 & 0.579 & 0.436 & 0.141 & 0.059 & 0.003 & 0.000  \\
\end{longtable}

\newpage

\setlength{\tabcolsep}{\tabcolsep}

\renewcommand{\arraystretch}{1.0}

\footnotesize
\begin{longtable}{rccccccccccc}
\caption{Level densities for $^{23}$Ne}
E $(MeV)$ & J=$\dfrac{1}{2}$ & J=$\dfrac{3}{2}$ & J=$\dfrac{5}{2}$ & J=$\dfrac{7}{2}$ & J=$\dfrac{9}{2}$ & J=$\dfrac{11}{2}$ & J=$\dfrac{13}{2}$ & J=$\dfrac{15}{2}$ & J=$\dfrac{17}{2}$ & J=$\dfrac{19}{2}$ & J=$\dfrac{21}{2}$ \\
\hline\\
\endhead
0.50 & 0.218 & 0.358 & 0.419 & 0.306 & 0.246 & 0.124 & 0.052 & 0.008 & 0.000 & 0.000 & 0.000 \\
1.50 & 0.287 & 0.479 & 0.576 & 0.465 & 0.351 & 0.185 & 0.080 & 0.013 & 0.002 & 0.000 & 0.000 \\
2.50 & 0.384 & 0.636 & 0.782 & 0.640 & 0.527 & 0.260 & 0.138 & 0.019 & 0.004 & 0.000 & 0.000 \\
3.50 & 0.488 & 0.866 & 1.006 & 0.884 & 0.725 & 0.416 & 0.199 & 0.051 & 0.007 & 0.000 & 0.000 \\
4.50 & 0.683 & 1.139 & 1.398 & 1.202 & 0.979 & 0.579 & 0.322 & 0.078 & 0.015 & 0.000 & 0.000 \\
5.50 & 0.890 & 1.497 & 1.815 & 1.601 & 1.318 & 0.796 & 0.444 & 0.145 & 0.032 & 0.000 & 0.000 \\
6.50 & 1.118 & 1.922 & 2.363 & 2.110 & 1.785 & 1.092 & 0.608 & 0.219 & 0.053 & 0.004 & 0.000 \\
7.50 & 1.396 & 2.400 & 2.963 & 2.729 & 2.293 & 1.495 & 0.873 & 0.316 & 0.109 & 0.010 & 0.000 \\
\end{longtable}

\newpage

\setlength{\tabcolsep}{\tabcolsep}

\renewcommand{\arraystretch}{1.0}

\footnotesize
\begin{longtable}{rccccccccccc}
\caption{Level densities for $^{24}$Ne}
E $(MeV)$ & J=0 & J=1 & J=2 & J=3 & J=4 & J=5 & J=6 & J=7 & J=8 & J=9 & J=10 \\
\hline\\
\endhead
0.50 & 0.151 & 0.107 & 0.271 & 0.152 & 0.176 & 0.068 & 0.036 & 0.006 & 0.000 & 0.000 & 0.000  \\
1.50 & 0.208 & 0.145 & 0.375 & 0.263 & 0.286 & 0.101 & 0.054 & 0.010 & 0.000 & 0.000 & 0.000  \\
2.50 & 0.257 & 0.225 & 0.512 & 0.352 & 0.392 & 0.184 & 0.084 & 0.017 & 0.000 & 0.000 & 0.000  \\
3.50 & 0.319 & 0.333 & 0.723 & 0.508 & 0.542 & 0.257 & 0.156 & 0.031 & 0.005 & 0.000 & 0.000  \\
4.50 & 0.390 & 0.440 & 0.943 & 0.706 & 0.707 & 0.351 & 0.222 & 0.045 & 0.010 & 0.000 & 0.000  \\
5.50 & 0.482 & 0.592 & 1.196 & 0.922 & 1.025 & 0.523 & 0.327 & 0.098 & 0.017 & 0.000 & 0.000  \\
6.50 & 0.612 & 0.838 & 1.590 & 1.260 & 1.322 & 0.744 & 0.481 & 0.151 & 0.045 & 0.000 & 0.000  \\
7.50 & 0.757 & 1.119 & 2.107 & 1.727 & 1.719 & 1.016 & 0.680 & 0.233 & 0.073 & 0.000 & 0.000  \\
\end{longtable}

\newpage

\setlength{\tabcolsep}{\tabcolsep}

\renewcommand{\arraystretch}{1.0}

\footnotesize
\begin{longtable}{rcccccccccc}
\caption{Level densities for $^{25}$Ne}
E $(MeV)$ & J=$\dfrac{1}{2}$ & J=$\dfrac{3}{2}$ & J=$\dfrac{5}{2}$ & J=$\dfrac{7}{2}$ & J=$\dfrac{9}{2}$ & J=$\dfrac{11}{2}$ & J=$\dfrac{13}{2}$ & J=$\dfrac{15}{2}$ & J=$\dfrac{17}{2}$ & J=$\dfrac{19}{2}$ \\
\hline\\
\endhead
0.50 & 0.219 & 0.372 & 0.446 & 0.348 & 0.251 & 0.096 & 0.030 & 0.000 & 0.000 & 0.000  \\
1.50 & 0.326 & 0.544 & 0.587 & 0.481 & 0.359 & 0.151 & 0.046 & 0.006 & 0.000 & 0.000  \\
2.50 & 0.416 & 0.698 & 0.836 & 0.630 & 0.478 & 0.216 & 0.086 & 0.011 & 0.003 & 0.000  \\
3.50 & 0.558 & 0.929 & 1.129 & 0.959 & 0.671 & 0.317 & 0.126 & 0.028 & 0.006 & 0.000  \\
4.50 & 0.724 & 1.204 & 1.479 & 1.242 & 0.972 & 0.471 & 0.204 & 0.046 & 0.010 & 0.000  \\
5.50 & 0.929 & 1.569 & 1.919 & 1.645 & 1.273 & 0.715 & 0.290 & 0.086 & 0.017 & 0.000  \\
6.50 & 1.180 & 2.002 & 2.479 & 2.130 & 1.718 & 0.972 & 0.465 & 0.138 & 0.035 & 0.000  \\
7.50 & 1.477 & 2.507 & 3.080 & 2.780 & 2.238 & 1.350 & 0.647 & 0.212 & 0.055 & 0.004  \\
\end{longtable}

\newpage

\setlength{\tabcolsep}{\tabcolsep}

\renewcommand{\arraystretch}{1.0}

\footnotesize
\begin{longtable}{rllllllllll}
\caption{Level densities for $^{26}$Ne}
E $(MeV)$ & J=0 & J=1 & J=2 & J=3 & J=4 & J=5 & J=6 & J=7 & J=8 & J=9   \\
\hline\\
\endhead
0.50 & 0.204 & 0.147 & 0.359 & 0.191 & 0.177 & 0.051 & 0.027 & 0.000 & 0.000 & 0.000  \\
1.50 & 0.262 & 0.206 & 0.506 & 0.303 & 0.282 & 0.075 & 0.044 & 0.004 & 0.000 & 0.000  \\
2.50 & 0.323 & 0.298 & 0.672 & 0.415 & 0.380 & 0.152 & 0.064 & 0.008 & 0.003 & 0.000  \\
3.50 & 0.398 & 0.399 & 0.883 & 0.587 & 0.576 & 0.221 & 0.100 & 0.014 & 0.005 & 0.000  \\
4.50 & 0.481 & 0.547 & 1.140 & 0.828 & 0.799 & 0.335 & 0.177 & 0.027 & 0.010 & 0.000  \\
5.50 & 0.602 & 0.755 & 1.460 & 1.078 & 1.051 & 0.520 & 0.282 & 0.043 & 0.018 & 0.000  \\
6.50 & 0.717 & 0.970 & 1.850 & 1.430 & 1.374 & 0.716 & 0.402 & 0.094 & 0.029 & 0.000  \\
7.50 & 0.847 & 1.269 & 2.316 & 1.919 & 1.762 & 0.967 & 0.581 & 0.146 & 0.044 & 0.000  \\
\end{longtable}

\newpage

\setlength{\tabcolsep}{\tabcolsep}

\renewcommand{\arraystretch}{1.0}

\footnotesize
\begin{longtable}{rlllllllll}
\caption{Level densities for $^{27}$Ne}
E $(MeV)$ & J=$\dfrac{1}{2}$ & J=$\dfrac{3}{2}$ & J=$\dfrac{5}{2}$ & J=$\dfrac{7}{2}$ & J=$\dfrac{9}{2}$ & J=$\dfrac{11}{2}$ & J=$\dfrac{13}{2}$ & J=$\dfrac{15}{2}$ & J=$\dfrac{17}{2}$  \\
\hline\\
\endhead
0.50 & 0.294 & 0.510 & 0.489 & 0.334 & 0.210 & 0.100 & 0.005 & 0.000 & 0.000  \\
1.50 & 0.385 & 0.686 & 0.700 & 0.495 & 0.309 & 0.145 & 0.025 & 0.000 & 0.000  \\
2.50 & 0.500 & 0.882 & 0.933 & 0.691 & 0.467 & 0.202 & 0.048 & 0.000 & 0.000  \\
3.50 & 0.637 & 1.159 & 1.195 & 0.945 & 0.642 & 0.274 & 0.084 & 0.006 & 0.000  \\
4.50 & 0.813 & 1.446 & 1.562 & 1.268 & 0.885 & 0.405 & 0.140 & 0.014 & 0.000  \\
5.50 & 1.003 & 1.787 & 1.952 & 1.664 & 1.196 & 0.606 & 0.212 & 0.029 & 0.000  \\
6.50 & 1.239 & 2.183 & 2.419 & 2.110 & 1.568 & 0.824 & 0.342 & 0.055 & 0.000  \\
7.50 & 1.508 & 2.635 & 2.989 & 2.657 & 1.987 & 1.122 & 0.516 & 0.094 & 0.000  \\
\end{longtable}

\setlength{\tabcolsep}{\tabcolsep}

\renewcommand{\arraystretch}{1.0}

\footnotesize
\begin{longtable}{rcccccccc}
\caption{Level densities for $^{28}$Ne}
E $(MeV)$ & J=0 & J=1 & J=2 & J=3 & J=4 & J=5 & J=6 & J=7  \\
\hline\\
\endhead
0.50 & 0.178 & 0.127 & 0.335 & 0.122 & 0.164 & 0.019 & 0.008 & 0.000  \\
1.50 & 0.227 & 0.185 & 0.452 & 0.193 & 0.245 & 0.030 & 0.019 & 0.000  \\
2.50 & 0.287 & 0.270 & 0.588 & 0.287 & 0.346 & 0.052 & 0.039 & 0.000  \\
3.50 & 0.361 & 0.369 & 0.754 & 0.425 & 0.477 & 0.102 & 0.074 & 0.000  \\
4.50 & 0.450 & 0.496 & 0.959 & 0.590 & 0.645 & 0.161 & 0.137 & 0.000  \\
5.50 & 0.551 & 0.635 & 1.207 & 0.807 & 0.855 & 0.257 & 0.216 & 0.000  \\
6.50 & 0.664 & 0.810 & 1.473 & 1.080 & 1.093 & 0.394 & 0.312 & 0.000  \\
7.50 & 0.773 & 1.012 & 1.778 & 1.372 & 1.367 & 0.570 & 0.424 & 0.014  \\
8.50 & 0.869 & 1.211 & 2.150 & 1.750 & 1.664 & 0.825 & 0.548 & 0.041  \\
9.50 & 0.954 & 1.425 & 2.511 & 2.134 & 1.989 & 1.129 & 0.680 & 0.097  \\
10.50 & 1.023 & 1.653 & 2.880 & 2.540 & 2.308 & 1.456 & 0.818 & 0.185 \\
11.50 & 1.088 & 1.895 & 3.278 & 2.969 & 2.659 & 1.800 & 0.948 & 0.287 \\
12.50 & 1.155 & 2.185 & 3.662 & 3.434 & 3.065 & 2.154 & 1.130 & 0.369 \\
13.50 & 1.231 & 2.490 & 4.065 & 3.908 & 3.535 & 2.498 & 1.334 & 0.441 \\
14.50 & 1.324 & 2.819 & 4.482 & 4.383 & 4.021 & 2.803 & 1.567 & 0.559 \\
15.50 & 1.415 & 3.163 & 4.893 & 4.845 & 4.512 & 3.044 & 1.799 & 0.800 \\
16.50 & 1.506 & 3.509 & 5.296 & 5.267 & 4.913 & 3.167 & 1.964 & 1.030 \\
17.50 & 1.589 & 3.841 & 5.687 & 5.624 & 5.213 & 3.207 & 2.070 & 1.019 \\
18.50 & 1.667 & 4.129 & 6.023 & 5.907 & 5.401 & 3.183 & 2.096 &   -   \\
19.50 & 1.734 & 4.359 & 6.300 & 6.106 & 5.500 & 3.173 & 2.083 &   -   \\
20.50 & 1.788 & 4.524 & 6.498 & 6.238 & 5.535 & 3.204 &   -   &   -   \\
\\
\end{longtable}

\newpage

\setlength{\tabcolsep}{\tabcolsep}

\renewcommand{\arraystretch}{1.0}

\footnotesize
\begin{longtable}{rlllccl}
\caption{Level densities for $^{29}$Ne}
E $(MeV)$ & J=$\dfrac{1}{2}$ & J=$\dfrac{3}{2}$ & J=$\dfrac{5}{2}$ & J=$\dfrac{7}{2}$ & J=$\dfrac{9}{2}$ & J=$\dfrac{11}{2}$ \\
\hline\\
\endhead
0.50 & 0.282 & 0.256 & 0.232 & 0.189 & 0.096 & 0.022  \\
1.50 & 0.388 & 0.312 & 0.334 & 0.271 & 0.157 & 0.075  \\
2.50 & 0.422 & 0.398 & 0.481 & 0.363 & 0.219 & 0.169  \\
3.50 & 0.389 & 0.502 & 0.674 & 0.481 & 0.275 & 0.251  \\
4.50 & 0.355 & 0.637 & 0.882 & 0.654 & 0.360 & 0.249  \\
5.50 & 0.358 & 0.796 & 1.047 & 0.848 & 0.496 & 0.186  \\
6.50 & 0.408 & 0.963 & 1.119 & 1.017 & 0.682 & 0.155  \\
7.50 & 0.488 & 1.106 & 1.105 & 1.136 & 0.866 & 0.221  \\
8.50 & 0.583 & 1.218 & 1.062 & 1.205 & 0.988 & 0.339  \\
9.50 & 0.666 & 1.321 & 1.082 & 1.255 & 1.019 & 0.418  \\
10.50 & 0.735 & 1.423 & 1.221 & 1.331 & 0.998 & 0.384 \\
11.50 & 0.772 & 1.548 & 1.485 & 1.412 & 0.971 & 0.300 \\
12.50 & 0.809 & 1.692 & 1.802 & 1.484 & 0.948 & 0.316 \\
13.50 & 0.867 & 1.791 & 2.069 & 1.527 & 0.907 & 0.499 \\
14.50 & 0.944 & 1.787 & 2.158 &   -   &   -   & 0.556 \\
\\
\end{longtable}

\newpage

\setlength{\tabcolsep}{\tabcolsep}

\renewcommand{\arraystretch}{1.0}

\footnotesize
\begin{longtable}{rllllllcl}
\caption{Level densities for $^{20}$Na}
E $(MeV)$ & J=0 & J=1 & J=2 & J=3 & J=4 & J=5 & J=6 & J=7 \\
\hline\\
\endhead
0.50 & 0.055 & 0.216 & 0.267 & 0.284 & 0.173 & 0.117 & 0.029 & 0.020  \\
1.50 & 0.085 & 0.297 & 0.372 & 0.406 & 0.259 & 0.177 & 0.055 & 0.047  \\
2.50 & 0.123 & 0.412 & 0.523 & 0.564 & 0.356 & 0.253 & 0.095 & 0.084  \\
3.50 & 0.176 & 0.539 & 0.705 & 0.746 & 0.483 & 0.367 & 0.165 & 0.134  \\
4.50 & 0.235 & 0.695 & 0.920 & 0.952 & 0.660 & 0.499 & 0.253 & 0.195  \\
5.50 & 0.309 & 0.865 & 1.172 & 1.196 & 0.859 & 0.659 & 0.354 & 0.250  \\
6.50 & 0.389 & 1.061 & 1.462 & 1.462 & 1.087 & 0.829 & 0.463 & 0.307  \\
7.50 & 0.476 & 1.274 & 1.778 & 1.754 & 1.345 & 1.023 & 0.580 & 0.353  \\
8.50 & 0.571 & 1.498 & 2.101 & 2.046 & 1.653 & 1.223 & 0.694 & 0.400  \\
9.50 & 0.665 & 1.735 & 2.447 & 2.356 & 1.968 & 1.437 & 0.814 & 0.455  \\
10.50 & 0.762 & 1.975 & 2.788 & 2.685 & 2.302 & 1.659 & 0.939 & 0.509 \\
11.50 & 0.853 & 2.213 & 3.120 & 3.012 & 2.627 & 1.873 & 1.058 & 0.547 \\
12.50 & 0.938 & 2.444 & 3.431 & 3.337 & 2.948 & 2.075 & 1.166 & 0.542 \\
13.50 & 1.013 & 2.657 & 3.707 & 3.652 & 3.221 & 2.251 & 1.253 & 0.502 \\
14.50 & 1.076 & 2.850 & 3.938 & 3.941 & 3.444 & 2.386 & 1.306 & 0.478 \\
15.50 & 1.125 & 3.021 & 4.136 & 4.186 & 3.608 & 2.465 & 1.316 & 0.524 \\
16.50 & 1.161 & 3.157 & 4.263 & 4.384 & 3.695 & 2.483 &   -   & 0.588 \\
\\
\end{longtable}

\setlength{\tabcolsep}{\tabcolsep}

\renewcommand{\arraystretch}{1.0}

\footnotesize
\begin{longtable}{rllllllccc}
\caption{Level densities for $^{21}$Na}
E $(MeV)$ & J=$\dfrac{1}{2}$ & J=$\dfrac{3}{2}$ & J=$\dfrac{5}{2}$ & J=$\dfrac{7}{2}$ & J=$\dfrac{9}{2}$ & J=$\dfrac{11}{2}$ & J=$\dfrac{13}{2}$ & J=$\dfrac{15}{2}$ & J=$\dfrac{17}{2}$ \\
\hline\\
\endhead
0.50 & 0.062 & 0.105 & 0.126 & 0.082 & 0.068 & 0.017 & 0.010 & 0.000 & 0.000  \\
1.50 & 0.101 & 0.147 & 0.175 & 0.130 & 0.102 & 0.030 & 0.017 & 0.000 & 0.000  \\
2.50 & 0.143 & 0.230 & 0.280 & 0.184 & 0.159 & 0.059 & 0.030 & 0.000 & 0.000  \\
3.50 & 0.192 & 0.310 & 0.393 & 0.304 & 0.238 & 0.098 & 0.049 & 0.005 & 0.000  \\
4.50 & 0.269 & 0.431 & 0.526 & 0.426 & 0.353 & 0.157 & 0.085 & 0.015 & 0.007  \\
5.50 & 0.372 & 0.622 & 0.752 & 0.617 & 0.485 & 0.259 & 0.133 & 0.028 & 0.016  \\
6.50 & 0.492 & 0.813 & 1.005 & 0.852 & 0.700 & 0.370 & 0.205 & 0.048 & 0.033  \\
7.50 & 0.643 & 1.075 & 1.329 & 1.123 & 0.955 & 0.546 & 0.313 & 0.086 & 0.068  \\
\end{longtable}

\newpage

\setlength{\tabcolsep}{\tabcolsep}

\renewcommand{\arraystretch}{1.0}

\footnotesize
\begin{longtable}{rllllllllccc}
\caption{Level densities for $^{22}$Na}
E $(MeV)$ & J=0 & J=1 & J=2 & J=3 & J=4 & J=5 & J=6 & J=7 & J=8 & J=9 & J=10  \\
\hline\\
\endhead
0.50 & 0.086 & 0.207 & 0.238 & 0.228 & 0.164 & 0.104 & 0.048 & 0.022 & 0.000 & 0.000 & 0.000  \\
1.50 & 0.111 & 0.279 & 0.334 & 0.375 & 0.288 & 0.165 & 0.072 & 0.030 & 0.000 & 0.000 & 0.000  \\
2.50 & 0.151 & 0.360 & 0.441 & 0.499 & 0.395 & 0.272 & 0.107 & 0.061 & 0.000 & 0.000 & 0.000  \\
3.50 & 0.210 & 0.551 & 0.695 & 0.700 & 0.532 & 0.383 & 0.198 & 0.085 & 0.012 & 0.004 & 0.000  \\
4.50 & 0.267 & 0.705 & 0.916 & 1.008 & 0.762 & 0.556 & 0.301 & 0.163 & 0.021 & 0.009 & 0.000  \\
5.50 & 0.358 & 0.961 & 1.180 & 1.305 & 1.096 & 0.814 & 0.449 & 0.231 & 0.035 & 0.018 & 0.000  \\
6.50 & 0.475 & 1.238 & 1.633 & 1.811 & 1.450 & 1.083 & 0.673 & 0.327 & 0.083 & 0.038 & 0.000  \\
7.50 & 0.596 & 1.558 & 2.116 & 2.306 & 2.009 & 1.421 & 0.913 & 0.472 & 0.140 & 0.066 & 0.000  \\
\end{longtable}

\newpage

\setlength{\tabcolsep}{\tabcolsep}

\renewcommand{\arraystretch}{1.0}

\footnotesize
\begin{longtable}{rlllllllcccc}
\caption{Level densities for $^{23}$Na}
E $(MeV)$ & J=$\dfrac{1}{2}$ & J=$\dfrac{3}{2}$ & J=$\dfrac{5}{2}$ & J=$\dfrac{7}{2}$ & J=$\dfrac{9}{2}$ & J=$\dfrac{11}{2}$ & J=$\dfrac{13}{2}$ & J=$\dfrac{15}{2}$ & J=$\dfrac{17}{2}$ & J=$\dfrac{19}{2}$ & J=$\dfrac{21}{2}$ \\
\hline\\
\endhead
0.50 & 0.116 & 0.194 & 0.199 & 0.116 & 0.092 & 0.049 & 0.032 & 0.003 & 0.000 & 0.000 & 0.000   \\
1.50 & 0.151 & 0.270 & 0.295 & 0.220 & 0.158 & 0.070 & 0.046 & 0.005 & 0.002 & 0.000 & 0.000   \\
2.50 & 0.198 & 0.351 & 0.411 & 0.350 & 0.276 & 0.140 & 0.066 & 0.018 & 0.004 & 0.000 & 0.000   \\
3.50 & 0.328 & 0.534 & 0.549 & 0.465 & 0.385 & 0.222 & 0.111 & 0.034 & 0.007 & 0.000 & 0.000   \\
4.50 & 0.423 & 0.726 & 0.859 & 0.730 & 0.512 & 0.310 & 0.192 & 0.049 & 0.015 & 0.000 & 0.000   \\
5.50 & 0.542 & 0.922 & 1.116 & 1.015 & 0.795 & 0.443 & 0.278 & 0.093 & 0.029 & 0.000 & 0.000   \\
6.50 & 0.770 & 1.337 & 1.512 & 1.326 & 1.120 & 0.676 & 0.379 & 0.141 & 0.059 & 0.000 & 0.000   \\
7.50 & 0.971 & 1.686 & 2.052 & 1.915 & 1.504 & 0.960 & 0.620 & 0.222 & 0.091 & 0.003 & 0.000   \\
\end{longtable}

\newpage

\setlength{\tabcolsep}{\tabcolsep}

\renewcommand{\arraystretch}{1.0}

\footnotesize
\begin{longtable}{rllllllllccccc}
\caption{Level densities for $^{24}$Na}
E $(MeV)$ & J=0 & J=1 & J=2 & J=3 & J=4 & J=5 & J=6 & J=7 & J=8 & J=9 & J=10 & J=11 & J=12 \\
\hline\\
\endhead
0.50 & 0.166 & 0.467 & 0.630 & 0.662 & 0.531 & 0.320 & 0.163 & 0.086 & 0.010 & 0.000 & 0.000 & 0.000 & 0.000  \\
1.50 & 0.229 & 0.649 & 0.831 & 0.862 & 0.742 & 0.524 & 0.258 & 0.122 & 0.024 & 0.002 & 0.000 & 0.000 & 0.000  \\
2.50 & 0.318 & 0.905 & 1.200 & 1.192 & 0.984 & 0.729 & 0.400 & 0.184 & 0.041 & 0.004 & 0.000 & 0.000 & 0.000  \\
3.50 & 0.428 & 1.182 & 1.626 & 1.682 & 1.340 & 0.976 & 0.572 & 0.298 & 0.074 & 0.011 & 0.000 & 0.000 & 0.000  \\
4.50 & 0.554 & 1.522 & 2.107 & 2.208 & 1.887 & 1.318 & 0.805 & 0.437 & 0.132 & 0.026 & 0.000 & 0.000 & 0.000  \\
5.50 & 0.711 & 1.975 & 2.748 & 2.905 & 2.536 & 1.912 & 1.175 & 0.616 & 0.205 & 0.054 & 0.003 & 0.000 & 0.000  \\
6.50 & 0.940 & 2.580 & 3.481 & 3.713 & 3.343 & 2.547 & 1.596 & 0.885 & 0.303 & 0.090 & 0.008 & 0.000 & 0.000  \\
7.50 & 1.170 & 3.226 & 4.559 & 4.959 & 4.279 & 3.301 & 2.148 & 1.227 & 0.431 & 0.141 & 0.016 & 0.000 & 0.000  \\
\end{longtable}

\newpage

\setlength{\tabcolsep}{\tabcolsep}

\renewcommand{\arraystretch}{1.0}

\footnotesize
\begin{longtable}{rlllllllccccc}
\caption{Level densities for $^{25}$Na}
E $(MeV)$ & J=$\dfrac{1}{2}$ & J=$\dfrac{3}{2}$ & J=$\dfrac{5}{2}$ & J=$\dfrac{7}{2}$ & J=$\dfrac{9}{2}$ & J=$\dfrac{11}{2}$ & J=$\dfrac{13}{2}$ & J=$\dfrac{15}{2}$ & J=$\dfrac{17}{2}$ & J=$\dfrac{19}{2}$ & J=$\dfrac{21}{2}$ & J=$\dfrac{23}{2}$ \\
\hline\\
\endhead
0.50 & 0.298 & 0.571 & 0.600 & 0.486 & 0.431 & 0.222 & 0.082 & 0.022 & 0.002 & 0.000 & 0.000 & 0.000  \\
1.50 & 0.403 & 0.765 & 0.892 & 0.694 & 0.592 & 0.314 & 0.155 & 0.032 & 0.007 & 0.000 & 0.000 & 0.000  \\
2.50 & 0.548 & 0.967 & 1.182 & 1.050 & 0.859 & 0.459 & 0.223 & 0.076 & 0.011 & 0.000 & 0.000 & 0.000  \\
3.50 & 0.745 & 1.381 & 1.519 & 1.366 & 1.195 & 0.639 & 0.333 & 0.112 & 0.028 & 0.000 & 0.000 & 0.000  \\
4.50 & 1.020 & 1.771 & 2.081 & 1.836 & 1.549 & 0.950 & 0.538 & 0.168 & 0.043 & 0.000 & 0.000 & 0.000  \\
5.50 & 1.293 & 2.341 & 2.837 & 2.441 & 2.149 & 1.268 & 0.735 & 0.266 & 0.083 & 0.000 & 0.000 & 0.000  \\
6.50 & 1.633 & 2.973 & 3.596 & 3.343 & 2.919 & 1.781 & 0.997 & 0.418 & 0.127 & 0.010 & 0.000 & 0.000  \\
7.50 & 2.121 & 3.821 & 4.615 & 4.260 & 3.727 & 2.462 & 1.405 & 0.591 & 0.218 & 0.021 & 0.003 & 0.000  \\
\end{longtable}

\newpage

\setlength{\tabcolsep}{\tabcolsep}

\renewcommand{\arraystretch}{1.0}

\footnotesize
\begin{longtable}{rlllllllllcccl}
\caption{Level densities for $^{26}$Na}
E $(MeV)$ & J=0 & J=1 & J=2 & J=3 & J=4 & J=5 & J=6 & J=7 & J=8 & J=9 & J=10 & J=11 & J=12 \\
\hline\\
\endhead
0.50 & 0.279 & 0.780 & 1.132 & 1.063 & 0.847 & 0.541 & 0.261 & 0.094 & 0.012 & 0.003 & 0.000 & 0.000 & 0.000  \\
1.50 & 0.396 & 1.064 & 1.477 & 1.479 & 1.255 & 0.862 & 0.362 & 0.147 & 0.029 & 0.005 & 0.000 & 0.000 & 0.000  \\
2.50 & 0.526 & 1.408 & 1.878 & 1.930 & 1.616 & 1.158 & 0.622 & 0.211 & 0.052 & 0.008 & 0.000 & 0.000 & 0.000  \\
3.50 & 0.662 & 1.808 & 2.535 & 2.548 & 2.124 & 1.548 & 0.842 & 0.360 & 0.082 & 0.018 & 0.000 & 0.000 & 0.000  \\
4.50 & 0.837 & 2.282 & 3.208 & 3.385 & 2.957 & 2.039 & 1.153 & 0.536 & 0.141 & 0.030 & 0.000 & 0.000 & 0.000  \\
5.50 & 1.067 & 2.926 & 4.127 & 4.282 & 3.788 & 2.799 & 1.604 & 0.773 & 0.253 & 0.055 & 0.004 & 0.000 & 0.000  \\
6.50 & 1.339 & 3.694 & 5.216 & 5.462 & 4.783 & 3.597 & 2.211 & 1.086 & 0.373 & 0.101 & 0.010 & 0.000 & 0.000  \\
7.50 & 1.676 & 4.640 & 6.558 & 6.893 & 6.147 & 4.682 & 2.877 & 1.539 & 0.568 & 0.162 & 0.018 & 0.000 & 0.000  \\
\end{longtable}

\newpage

\setlength{\tabcolsep}{\tabcolsep}

\renewcommand{\arraystretch}{1.0}

\footnotesize
\begin{longtable}{rllllllcccccc}
\caption{Level densities for $^{27}$Na}
E $(MeV)$ & J=$\dfrac{1}{2}$ & J=$\dfrac{3}{2}$ & J=$\dfrac{5}{2}$ & J=$\dfrac{7}{2}$ & J=$\dfrac{9}{2}$ & J=$\dfrac{11}{2}$ & J=$\dfrac{13}{2}$ & J=$\dfrac{15}{2}$ & J=$\dfrac{17}{2}$ & J=$\dfrac{19}{2}$ & J=$\dfrac{21}{2}$ & J=$\dfrac{23}{2}$ \\
\hline\\
\endhead
0.50 & 0.383 & 0.680 & 0.791 & 0.590 & 0.440 & 0.197 & 0.080 & 0.008 & 0.003 & 0.000 & 0.000 & 0.000  \\
1.50 & 0.550 & 0.940 & 1.111 & 0.835 & 0.671 & 0.279 & 0.130 & 0.014 & 0.005 & 0.000 & 0.000 & 0.000  \\
2.50 & 0.746 & 1.221 & 1.423 & 1.189 & 0.948 & 0.458 & 0.184 & 0.038 & 0.009 & 0.000 & 0.000 & 0.000  \\
3.50 & 0.952 & 1.634 & 1.857 & 1.551 & 1.253 & 0.674 & 0.311 & 0.062 & 0.013 & 0.000 & 0.000 & 0.000  \\
4.50 & 1.199 & 2.113 & 2.478 & 2.126 & 1.656 & 0.913 & 0.470 & 0.111 & 0.024 & 0.000 & 0.000 & 0.000  \\
5.50 & 1.572 & 2.712 & 3.241 & 2.870 & 2.251 & 1.240 & 0.646 & 0.190 & 0.059 & 0.000 & 0.000 & 0.000  \\
6.50 & 1.965 & 3.486 & 4.038 & 3.661 & 3.001 & 1.733 & 0.896 & 0.303 & 0.093 & 0.000 & 0.000 & 0.000  \\
7.50 & 2.474 & 4.289 & 5.183 & 4.612 & 3.806 & 2.397 & 1.275 & 0.439 & 0.140 & 0.010 & 0.000 & 0.000  \\
\end{longtable}

\newpage

\setlength{\tabcolsep}{\tabcolsep}

\renewcommand{\arraystretch}{1.0}

\footnotesize
\begin{longtable}{rllllllccccc}
\caption{Level densities for $^{28}$Na}
E $(MeV)$ & J=0 & J=1 & J=2 & J=3 & J=4 & J=5 & J=6 & J=7 & J=8 & J=9 & J=10  \\
\hline\\
\endhead
0.50 & 0.319 & 0.764 & 0.999 & 0.976 & 0.696 & 0.397 & 0.162 & 0.043 & 0.000 & 0.000 & 0.000  \\
1.50 & 0.421 & 1.027 & 1.283 & 1.270 & 0.962 & 0.569 & 0.263 & 0.070 & 0.000 & 0.000 & 0.000  \\
2.50 & 0.543 & 1.325 & 1.760 & 1.753 & 1.339 & 0.793 & 0.393 & 0.108 & 0.008 & 0.000 & 0.000  \\
3.50 & 0.669 & 1.676 & 2.224 & 2.262 & 1.803 & 1.117 & 0.563 & 0.168 & 0.015 & 0.000 & 0.000  \\
4.50 & 0.845 & 2.110 & 2.816 & 2.865 & 2.319 & 1.513 & 0.808 & 0.285 & 0.027 & 0.004 & 0.000  \\
5.50 & 1.039 & 2.624 & 3.499 & 3.655 & 2.968 & 1.960 & 1.114 & 0.435 & 0.057 & 0.011 & 0.000  \\
6.50 & 1.245 & 3.259 & 4.382 & 4.620 & 3.774 & 2.550 & 1.484 & 0.622 & 0.113 & 0.025 & 0.000  \\
7.50 & 1.523 & 3.957 & 5.311 & 5.636 & 4.716 & 3.309 & 1.941 & 0.852 & 0.211 & 0.054 & 0.000  \\
\end{longtable}

\setlength{\tabcolsep}{\tabcolsep}

\renewcommand{\arraystretch}{1.0}

\footnotesize
\begin{longtable}{rllllllccc}
\caption{Level densities for $^{29}$Na}
E $(MeV)$ & J=$\dfrac{1}{2}$ & J=$\dfrac{3}{2}$ & J=$\dfrac{5}{2}$ & J=$\dfrac{7}{2}$ & J=$\dfrac{9}{2}$ & J=$\dfrac{11}{2}$ & J=$\dfrac{13}{2}$ & J=$\dfrac{15}{2}$ & J=$\dfrac{17}{2}$  \\
\hline\\
\endhead
0.50 & 0.328 & 0.509 & 0.552 & 0.403 & 0.275 & 0.092 & 0.029 & 0.000 & 0.000  \\
1.50 & 0.433 & 0.703 & 0.741 & 0.566 & 0.390 & 0.148 & 0.056 & 0.000 & 0.000  \\
2.50 & 0.574 & 0.921 & 1.019 & 0.804 & 0.542 & 0.223 & 0.097 & 0.000 & 0.000  \\
3.50 & 0.731 & 1.167 & 1.350 & 1.088 & 0.766 & 0.323 & 0.158 & 0.000 & 0.000  \\
4.50 & 0.911 & 1.477 & 1.756 & 1.416 & 1.042 & 0.454 & 0.240 & 0.000 & 0.000  \\
5.50 & 1.146 & 1.875 & 2.213 & 1.851 & 1.359 & 0.658 & 0.343 & 0.012 & 0.000  \\
6.50 & 1.401 & 2.302 & 2.724 & 2.335 & 1.793 & 0.908 & 0.477 & 0.032 & 0.006  \\
7.50 & 1.695 & 2.781 & 3.325 & 2.877 & 2.276 & 1.234 & 0.628 & 0.076 & 0.021  \\
\end{longtable}

\newpage

\setlength{\tabcolsep}{\tabcolsep}

\renewcommand{\arraystretch}{1.0}

\footnotesize
\begin{longtable}{rllllcccc}
\caption{Level densities for $^{30}$Na}
E $(MeV)$ & J=0 & J=1 & J=2 & J=3 & J=4 & J=5 & J=6 & J=7 \\
\hline\\
\endhead
0.50 & 0.252 & 0.376 & 0.548 & 0.484 & 0.310 & 0.119 & 0.062 & 0.000  \\
1.50 & 0.304 & 0.491 & 0.702 & 0.628 & 0.416 & 0.189 & 0.114 & 0.000  \\
2.50 & 0.343 & 0.627 & 0.887 & 0.807 & 0.548 & 0.307 & 0.167 & 0.000  \\
3.50 & 0.373 & 0.771 & 1.072 & 1.004 & 0.713 & 0.452 & 0.217 & 0.000  \\
4.50 & 0.389 & 0.957 & 1.274 & 1.249 & 0.912 & 0.622 & 0.250 & 0.000  \\
5.50 & 0.398 & 1.141 & 1.503 & 1.504 & 1.156 & 0.794 & 0.288 & 0.006  \\
6.50 & 0.405 & 1.328 & 1.738 & 1.803 & 1.411 & 0.971 & 0.367 & 0.028  \\
7.50 & 0.425 & 1.520 & 1.976 & 2.101 & 1.668 & 1.120 & 0.516 & 0.132  \\
8.50 & 0.456 & 1.721 & 2.220 & 2.399 & 1.937 & 1.282 & 0.688 & 0.414  \\
9.50 & 0.509 & 1.941 & 2.488 & 2.696 & 2.217 & 1.461 & 0.844 & 0.673  \\
10.50 & 0.587 & 2.159 & 2.753 & 2.979 & 2.511 & 1.701 & 0.950 & 0.492 \\
11.50 & 0.687 & 2.369 & 3.025 & 3.274 & 2.809 & 1.968 & 1.020 & 0.215 \\
12.50 & 0.805 & 2.557 & 3.317 & 3.536 & 3.103 & 2.212 & 1.043 & 0.158 \\
13.50 & 0.923 & 2.705 & 3.586 & 3.772 & 3.328 & 2.397 & 1.068 & 0.294 \\
14.50 & 1.023 & 2.817 & 3.809 & 3.965 & 3.468 & 2.466 & 1.147 & 0.484 \\
15.50 & 1.095 & 2.867 & 3.973 & 4.079 & 3.497 &  -    & 1.255 & 0.612 \\
16.50 & 1.138 & 2.883 & 4.082 & 4.130 &   -   &  -    &  -    &   -   \\
17.50 & 1.162 & 2.875 & 4.122 & 4.144 &   -   &  -    &  -    &   -   \\
\\
\end{longtable}

\setlength{\tabcolsep}{\tabcolsep}

\renewcommand{\arraystretch}{1.0}

\footnotesize
\begin{longtable}{rlllll}
\caption{Level densities for $^{31}$Na}
E $(MeV)$ & J=$\dfrac{1}{2}$ & J=$\dfrac{3}{2}$ & J=$\dfrac{5}{2}$ & J=$\dfrac{7}{2}$ & J=$\dfrac{9}{2}$ \\
\hline\\
\endhead
0.50 & 0.058 & 0.252 & 0.222 & 0.006 & 0.158 \\
1.50 & 0.087 & 0.296 & 0.223 & 0.023 & 0.227 \\
2.50 & 0.116 & 0.320 & 0.214 & 0.065 & 0.249 \\
3.50 & 0.138 & 0.334 & 0.216 & 0.134 & 0.216 \\
4.50 & 0.146 & 0.348 & 0.272 & 0.215 & 0.184 \\
5.50 & 0.149 & 0.365 & 0.369 & 0.270 & 0.201 \\
6.50 & 0.163 & 0.381 & 0.505 & 0.305 & 0.269 \\
7.50 & 0.213 & 0.384 & 0.647 & 0.356 & 0.334 \\
8.50 & 0.290 & 0.383 & 0.732 & 0.446 & 0.375 \\
\\
\end{longtable}

\newpage

\setlength{\tabcolsep}{\tabcolsep}

\renewcommand{\arraystretch}{1.0}

\footnotesize
\begin{longtable}{rlllllllcccc}
\caption{Level densities for $^{22}$Mg}
E $(MeV)$ & J=0 & J=1 & J=2 & J=3 & J=4 & J=5 & J=6 & J=7 & J=8 & J=9 & J=10  \\
\hline\\
\endhead
0.50 & 0.083 & 0.043 & 0.137 & 0.053 & 0.087 & 0.025 & 0.015 & 0.003 & 0.000 & 0.000 & 0.000  \\
1.50 & 0.108 & 0.067 & 0.199 & 0.090 & 0.137 & 0.038 & 0.037 & 0.005 & 0.000 & 0.000 & 0.000  \\
2.50 & 0.139 & 0.121 & 0.294 & 0.144 & 0.200 & 0.056 & 0.061 & 0.008 & 0.002 & 0.000 & 0.000  \\
3.50 & 0.175 & 0.163 & 0.390 & 0.238 & 0.309 & 0.098 & 0.088 & 0.022 & 0.004 & 0.000 & 0.000  \\
4.50 & 0.234 & 0.241 & 0.545 & 0.327 & 0.421 & 0.151 & 0.138 & 0.033 & 0.008 & 0.000 & 0.000  \\
5.50 & 0.313 & 0.357 & 0.751 & 0.518 & 0.581 & 0.245 & 0.216 & 0.054 & 0.019 & 0.000 & 0.000  \\
6.50 & 0.405 & 0.478 & 0.986 & 0.722 & 0.836 & 0.372 & 0.319 & 0.084 & 0.037 & 0.000 & 0.000  \\
7.50 & 0.501 & 0.663 & 1.274 & 0.979 & 1.125 & 0.579 & 0.436 & 0.141 & 0.059 & 0.003 & 0.000  \\
\end{longtable}

\newpage

\setlength{\tabcolsep}{\tabcolsep}

\renewcommand{\arraystretch}{1.0}

\footnotesize
\begin{longtable}{rllllllllcccc}
\caption{Level densities for $^{23}$Mg}
E $(MeV)$ & J=$\dfrac{1}{2}$ & J=$\dfrac{3}{2}$ & J=$\dfrac{5}{2}$ & J=$\dfrac{7}{2}$ & J=$\dfrac{9}{2}$ & J=$\dfrac{11}{2}$ & J=$\dfrac{13}{2}$ & J=$\dfrac{15}{2}$ & J=$\dfrac{17}{2}$ & J=$\dfrac{19}{2}$ & J=$\dfrac{21}{2}$ & J=$\dfrac{23}{2}$ \\
\hline\\
\endhead
0.50 & 0.116 & 0.194 & 0.199 & 0.116 & 0.092 & 0.049 & 0.032 & 0.003 & 0.000 & 0.000 & 0.000 & 0.000  \\
1.50 & 0.151 & 0.270 & 0.295 & 0.220 & 0.158 & 0.070 & 0.046 & 0.005 & 0.002 & 0.000 & 0.000 & 0.000  \\
2.50 & 0.198 & 0.351 & 0.411 & 0.350 & 0.276 & 0.140 & 0.066 & 0.018 & 0.004 & 0.000 & 0.000 & 0.000  \\
3.50 & 0.328 & 0.534 & 0.549 & 0.465 & 0.385 & 0.222 & 0.111 & 0.034 & 0.007 & 0.000 & 0.000 & 0.000  \\
4.50 & 0.423 & 0.726 & 0.859 & 0.730 & 0.512 & 0.310 & 0.192 & 0.049 & 0.015 & 0.000 & 0.000 & 0.000  \\
5.50 & 0.542 & 0.922 & 1.116 & 1.015 & 0.795 & 0.443 & 0.278 & 0.093 & 0.029 & 0.000 & 0.000 & 0.000  \\
6.50 & 0.770 & 1.337 & 1.512 & 1.326 & 1.120 & 0.676 & 0.379 & 0.141 & 0.059 & 0.000 & 0.000 & 0.000  \\
7.50 & 0.971 & 1.686 & 2.052 & 1.915 & 1.504 & 0.960 & 0.620 & 0.222 & 0.091 & 0.003 & 0.000 & 0.000  \\
\end{longtable}

\newpage

\setlength{\tabcolsep}{\tabcolsep}

\renewcommand{\arraystretch}{1.0}

\footnotesize
\begin{longtable}{rllllllccccccc}
\caption{Level densities for $^{24}$Mg}
E $(MeV)$ & J=0 & J=1 & J=2 & J=3 & J=4 & J=5 & J=6 & J=7 & J=8 & J=9 & J=10 & J=11 & J=12  \\
\hline\\
\endhead
0.50 & 0.056 & 0.024 & 0.057 & 0.036 & 0.047 & 0.008 & 0.006 & 0.000 & 0.002 & 0.000 & 0.000 & 0.000 & 0.000  \\
1.50 & 0.070 & 0.032 & 0.086 & 0.049 & 0.062 & 0.011 & 0.009 & 0.000 & 0.003 & 0.000 & 0.000 & 0.000 & 0.000  \\
2.50 & 0.087 & 0.061 & 0.138 & 0.065 & 0.082 & 0.041 & 0.028 & 0.000 & 0.004 & 0.000 & 0.000 & 0.000 & 0.000  \\
3.50 & 0.116 & 0.083 & 0.226 & 0.119 & 0.122 & 0.056 & 0.039 & 0.002 & 0.007 & 0.000 & 0.000 & 0.000 & 0.000  \\
4.50 & 0.143 & 0.159 & 0.317 & 0.225 & 0.252 & 0.075 & 0.060 & 0.014 & 0.015 & 0.000 & 0.000 & 0.000 & 0.000  \\
5.50 & 0.191 & 0.208 & 0.414 & 0.296 & 0.331 & 0.184 & 0.118 & 0.021 & 0.021 & 0.000 & 0.000 & 0.000 & 0.000  \\
6.50 & 0.284 & 0.382 & 0.683 & 0.421 & 0.468 & 0.248 & 0.190 & 0.033 & 0.029 & 0.000 & 0.000 & 0.000 & 0.000  \\
7.50 & 0.350 & 0.492 & 0.872 & 0.741 & 0.675 & 0.354 & 0.273 & 0.088 & 0.056 & 0.000 & 0.000 & 0.000 & 0.000  \\
\end{longtable}

\newpage

\setlength{\tabcolsep}{\tabcolsep}

\renewcommand{\arraystretch}{1.0}

\footnotesize
\begin{longtable}{rllllllccccccc}
\caption{Level densities for $^{25}$Mg}
E $(MeV)$ & J=$\dfrac{1}{2}$ & J=$\dfrac{3}{2}$ & J=$\dfrac{5}{2}$ & J=$\dfrac{7}{2}$ & J=$\dfrac{9}{2}$ & J=$\dfrac{11}{2}$ & J=$\dfrac{13}{2}$ & J=$\dfrac{15}{2}$ & J=$\dfrac{17}{2}$ & J=$\dfrac{19}{2}$ & J=$\dfrac{21}{2}$ & J=$\dfrac{23}{2}$ & J=$\dfrac{25}{2}$ \\
\hline\\
\endhead
0.50 & 0.142 & 0.232 & 0.297 & 0.262 & 0.196 & 0.118 & 0.057 & 0.009 & 0.002 & 0.000 & 0.000 & 0.000 & 0.000  \\
1.50 & 0.245 & 0.422 & 0.526 & 0.399 & 0.294 & 0.167 & 0.094 & 0.021 & 0.005 & 0.000 & 0.000 & 0.000 & 0.000  \\
2.50 & 0.320 & 0.542 & 0.683 & 0.620 & 0.471 & 0.241 & 0.132 & 0.040 & 0.011 & 0.002 & 0.000 & 0.000 & 0.000  \\
3.50 & 0.451 & 0.794 & 0.896 & 0.809 & 0.685 & 0.433 & 0.216 & 0.061 & 0.018 & 0.003 & 0.000 & 0.000 & 0.000  \\
4.50 & 0.647 & 1.126 & 1.381 & 1.098 & 0.886 & 0.584 & 0.354 & 0.094 & 0.031 & 0.005 & 0.000 & 0.000 & 0.000  \\
5.50 & 0.870 & 1.485 & 1.824 & 1.650 & 1.284 & 0.776 & 0.476 & 0.189 & 0.064 & 0.007 & 0.000 & 0.000 & 0.000  \\
6.50 & 1.170 & 2.012 & 2.363 & 2.220 & 1.884 & 1.224 & 0.655 & 0.271 & 0.093 & 0.011 & 0.000 & 0.000 & 0.000  \\
7.50 & 1.490 & 2.615 & 3.154 & 2.993 & 2.417 & 1.676 & 1.009 & 0.375 & 0.149 & 0.027 & 0.000 & 0.000 & 0.000  \\
\end{longtable}

\newpage

\setlength{\tabcolsep}{\tabcolsep}

\renewcommand{\arraystretch}{1.0}

\footnotesize
\begin{longtable}{rllllllccccccc}
\caption{Level densities for $^{26}$Mg}
E $(MeV)$ & J=0 & J=1 & J=2 & J=3 & J=4 & J=5 & J=6 & J=7 & J=8 & J=9 & J=10 & J=11 & J=12  \\
\hline\\
\endhead
0.50 & 0.141 & 0.130 & 0.283 & 0.212 & 0.193 & 0.084 & 0.045 & 0.011 & 0.004 & 0.000 & 0.000 & 0.000 & 0.000   \\
1.50 & 0.180 & 0.185 & 0.384 & 0.301 & 0.318 & 0.171 & 0.078 & 0.020 & 0.006 & 0.000 & 0.000 & 0.000 & 0.000   \\
2.50 & 0.222 & 0.242 & 0.549 & 0.391 & 0.418 & 0.230 & 0.152 & 0.037 & 0.008 & 0.000 & 0.000 & 0.000 & 0.000   \\
3.50 & 0.308 & 0.410 & 0.807 & 0.582 & 0.625 & 0.307 & 0.204 & 0.081 & 0.018 & 0.000 & 0.000 & 0.000 & 0.000   \\
4.50 & 0.384 & 0.547 & 1.056 & 0.888 & 0.955 & 0.449 & 0.307 & 0.112 & 0.038 & 0.000 & 0.000 & 0.000 & 0.000   \\
5.50 & 0.508 & 0.780 & 1.376 & 1.194 & 1.231 & 0.768 & 0.480 & 0.165 & 0.058 & 0.007 & 0.001 & 0.000 & 0.000   \\
6.50 & 0.659 & 1.079 & 1.879 & 1.695 & 1.586 & 1.021 & 0.706 & 0.255 & 0.102 & 0.011 & 0.002 & 0.000 & 0.000   \\
7.50 & 0.845 & 1.497 & 2.564 & 2.251 & 2.210 & 1.402 & 0.938 & 0.427 & 0.166 & 0.017 & 0.004 & 0.000 & 0.000   \\
\end{longtable}

\newpage

\setlength{\tabcolsep}{\tabcolsep}

\renewcommand{\arraystretch}{1.0}

\footnotesize
\begin{longtable}{rlllllllcccccc}
\caption{Level densities for $^{27}$Mg}
E $(MeV)$ & J=$\dfrac{1}{2}$ & J=$\dfrac{3}{2}$ & J=$\dfrac{5}{2}$ & J=$\dfrac{7}{2}$ & J=$\dfrac{9}{2}$ & J=$\dfrac{11}{2}$ & J=$\dfrac{13}{2}$ & J=$\dfrac{15}{2}$ & J=$\dfrac{17}{2}$ & J=$\dfrac{19}{2}$ & J=$\dfrac{21}{2}$ & J=$\dfrac{23}{2}$ & J=$\dfrac{25}{2}$  \\
\hline\\
\endhead
0.50 & 0.316 & 0.467 & 0.544 & 0.477 & 0.369 & 0.201 & 0.077 & 0.011 & 0.004 & 0.000 & 0.000 & 0.000 & 0.000  \\
1.50 & 0.421 & 0.695 & 0.821 & 0.645 & 0.497 & 0.271 & 0.133 & 0.032 & 0.005 & 0.000 & 0.000 & 0.000 & 0.000  \\
2.50 & 0.593 & 0.984 & 1.172 & 1.000 & 0.754 & 0.374 & 0.186 & 0.046 & 0.014 & 0.001 & 0.000 & 0.000 & 0.000  \\
3.50 & 0.771 & 1.291 & 1.535 & 1.387 & 1.114 & 0.674 & 0.260 & 0.082 & 0.021 & 0.002 & 0.000 & 0.000 & 0.000  \\
4.50 & 0.979 & 1.655 & 2.006 & 1.826 & 1.479 & 0.899 & 0.484 & 0.120 & 0.036 & 0.003 & 0.000 & 0.000 & 0.000  \\
5.50 & 1.312 & 2.253 & 2.723 & 2.352 & 1.922 & 1.222 & 0.660 & 0.246 & 0.051 & 0.005 & 0.000 & 0.000 & 0.000  \\
6.50 & 1.753 & 2.997 & 3.606 & 3.425 & 2.785 & 1.604 & 0.897 & 0.359 & 0.120 & 0.010 & 0.000 & 0.000 & 0.000  \\
7.50 & 2.201 & 3.806 & 4.612 & 4.350 & 3.602 & 2.417 & 1.304 & 0.504 & 0.172 & 0.026 & 0.000 & 0.000 & 0.000  \\
\end{longtable}

\newpage

\setlength{\tabcolsep}{\tabcolsep}

\renewcommand{\arraystretch}{1.0}

\footnotesize
\begin{longtable}{rllllllllccccc}
\caption{Level densities for $^{28}$Mg}
E $(MeV)$ & J=0 & J=1 & J=2 & J=3 & J=4 & J=5 & J=6 & J=7 & J=8 & J=9 & J=10 & J=11 & J=12 \\
\hline\\
\endhead
0.50 & 0.196 & 0.219 & 0.405 & 0.279 & 0.285 & 0.127 & 0.045 & 0.008 & 0.003 & 0.000 & 0.000 & 0.000 & 0.000  \\
1.50 & 0.262 & 0.285 & 0.552 & 0.371 & 0.378 & 0.179 & 0.102 & 0.011 & 0.004 & 0.000 & 0.000 & 0.000 & 0.000  \\
2.50 & 0.339 & 0.420 & 0.730 & 0.516 & 0.543 & 0.251 & 0.145 & 0.031 & 0.008 & 0.000 & 0.000 & 0.000 & 0.000  \\
3.50 & 0.423 & 0.605 & 1.073 & 0.793 & 0.864 & 0.360 & 0.218 & 0.057 & 0.016 & 0.000 & 0.000 & 0.000 & 0.000  \\
4.50 & 0.516 & 0.809 & 1.393 & 1.115 & 1.125 & 0.604 & 0.350 & 0.083 & 0.023 & 0.002 & 0.000 & 0.000 & 0.000  \\
5.50 & 0.680 & 1.087 & 1.861 & 1.508 & 1.446 & 0.832 & 0.526 & 0.154 & 0.051 & 0.004 & 0.000 & 0.000 & 0.000  \\
6.50 & 0.837 & 1.411 & 2.397 & 2.053 & 1.975 & 1.140 & 0.714 & 0.245 & 0.091 & 0.009 & 0.000 & 0.000 & 0.000  \\
7.50 & 1.066 & 1.938 & 3.171 & 2.879 & 2.799 & 1.658 & 0.970 & 0.348 & 0.135 & 0.016 & 0.000 & 0.000 & 0.000  \\
\end{longtable}%

\end{document}